\documentclass[11pt]{article}
\pdfoutput=1
\usepackage{jheppub}
\usepackage{amsmath,amssymb,amsfonts,graphicx,color,mathtools}
\usepackage{arydshln}
\usepackage{hyperref}
\usepackage{subfig}
\usepackage[utf8]{inputenc}
\usepackage[titletoc]{appendix}
\allowdisplaybreaks

\title{Knitting Wormholes by Entanglement in Supergravity}

\author[a, b]{Vijay Balasubramanian}
\author[a]{\!, Matthew DeCross}
\author[a, c]{\!, G\'{a}bor S\'{a}rosi}

\affiliation[\,a]{David Rittenhouse Laboratory, University of Pennsylvania,\\
209 S.33rd Street, Philadelphia PA, 19104, U.S.A.}
\affiliation[\,b]{Theoretische Natuurkunde, Vrije Universiteit Brussel (VUB), and \\ International Solvay Institutes, Pleinlaan 2, B-1050 Brussels, Belgium.}
\affiliation[\,c]{CERN, Theoretical Physics Department, 1211 Geneva 23, Switzerland}

\emailAdd{vijay@physics.upenn.edu}
\emailAdd{mdecross@sas.upenn.edu}
\emailAdd{gabor.sarosi@cern.ch}

\abstract{
We construct a single-boundary wormhole geometry in type IIB supergravity by perturbing two stacks of $N$ extremal D3-branes in the decoupling limit. The solution interpolates from a two-sided planar AdS-Schwarzschild geometry in the interior, through a harmonic two-center solution in the intermediate region, to an asymptotic AdS space.  The construction involves a CPT twist in the gluing of the wormhole to the exterior throats that gives a global monodromy to some coordinates, while preserving orientability. The geometry has a dual interpretation in $\mathcal{N}=4$ $SU(2N)$ Super Yang-Mills theory in terms of a Higgsed $SU(2N) \to S(U(N) \times U(N))$ theory in which $\mathcal{O} (N^2)$ degrees of freedom in each $SU(N)$ sector are entangled in an approximate thermofield double state at a temperature much colder than the Higgs scale. We argue that the solution can be made long-lived by appropriate choice of parameters, and comment on mechanisms for generating traversability. We also describe a construction of a double wormhole between two universes.
}


\begin{document}
\begin{flushright}
\hfill{\tt CERN-TH-2020-153}
\end{flushright}

\maketitle
\parskip=10pt

\section{Introduction}

In the AdS/CFT correspondence, the entanglement structure of the boundary CFT encodes the geometry and topology of the bulk AdS space \cite{vanRaams, Maldacena_2003, Balasubramanian_2014, Marolf_2015, doi:10.1002/prop.201300020}, albeit in a complicated and nonlocal way.  We consider the dual of a state in the $\mathcal{N} = 4$ $SU(2N)$ Super Yang-Mills (SYM) theory in a Coulomb phase where the infrared modes are thermally entangled.  We argue that this system should correspond to an asymptotically AdS$_5\times S^5$ geometry with a single boundary and a long-lived interior wormhole.  The purpose of this paper is to construct this wormhole and describe its properties.

\begin{center}
\begin{figure}[hbtp!]
\includegraphics[width=\textwidth]{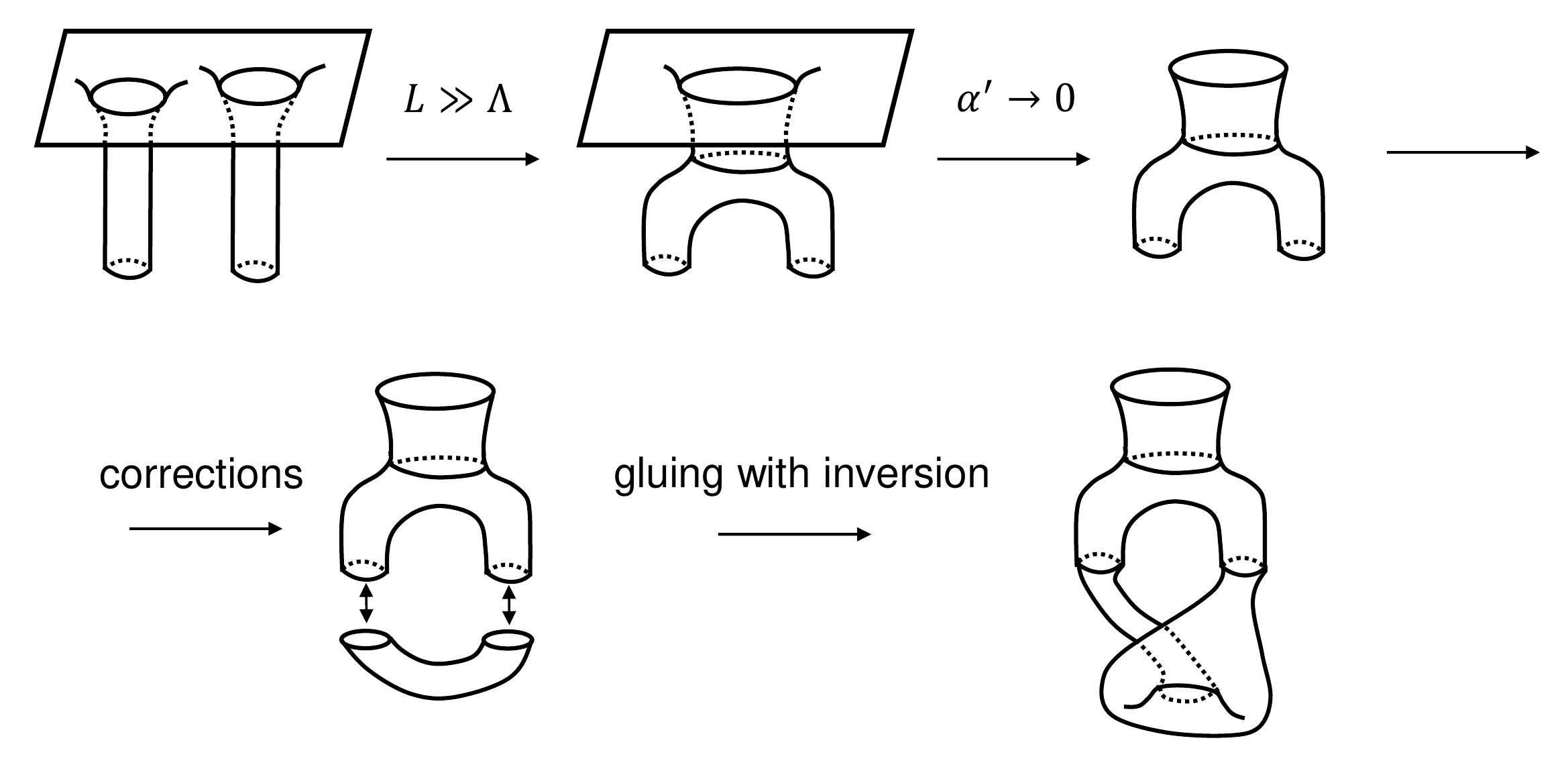}
\caption{Perturbative construction of an AdS wormhole solution in a single asymptotic space by taking the decoupling limit of two stacks of D3-branes and correcting for the finite temperature caused by the thermofield double entanglement structure below the scale $\Lambda$. Figure inspired by the near-horizon limit as depicted in \cite{costa, Michelson_1999}.\label{fig:wormhole1}}
\end{figure}
\end{center}

The simplest connection between wormholes and entanglement involves two copies of a CFT  entangled in the ``thermofield double" (TFD) state, a two-party purification of the thermal state on each factor:
\begin{align}
    |\text{TFD}\rangle = \sum_i e^{-\beta E_i / 2} |i\rangle \otimes |i^{\ast}\rangle \, .
\end{align}
where $|i^{\ast}\rangle$ indicates the CPT conjugate of $|i\rangle$. This system is dual to the eternal black hole, i.e., a wormhole between two asymptotically AdS universes \cite{Maldacena_2003}. To construct a wormhole between distant regions of a \emph{single} universe, we will study a state in the Coulomb branch of the $\mathcal{N}=4$ SYM theory \cite{KLEBANOV199989, PhysRevD.62.086003}, where the $SU(2N)$ gauge symmetry has been partially broken down to $S(U(N) \times U(N))$. In type IIB supergravity, the low energy effective theory dual to $\mathcal{N}=4$ SYM, this configuration corresponds to a multicenter solution sourced by two stacks of $N$ D3-branes \cite{CoulombBranch, PhysRevD.60.127902}. The procedure for constructing this geometry is illustrated in Fig.~\ref{fig:wormhole1}. We will begin with a two-centered BPS \cite{DUFF1991409} harmonic function solution as originally found in \cite{HOROWITZ1991197}, corresponding to two stacks of $N$ extremal D3-branes in (9+1)D Minkowski space and controlled by a parameter $L$. These stacks of D3-branes are separated by a distance $\Lambda$, corresponding to the Higgs scale. In the limit that $L\gg \Lambda$, there will be an AdS$_5 \times S^5$ geometry outside the region containing the two stacks of branes, which splits into two smaller AdS$_5 \times S^5$ regions as one nears either stack. Taking the limit $\alpha^{\prime} \to 0$ with the ratios of the five-sphere coordinates to $\alpha^{\prime}$ held fixed decouples the AdS regions from the asymptotically flat space, leaving a geometry which is asymptotically AdS \cite{largeN}. 

Now we heat up the solution by entangling the degrees of freedom living on each brane (in each $SU(N)$ sector of the Higgsed SYM) up to the Higgs scale $\Lambda$ in an approximate thermofield double state. This has the effect that in the IR of the field theory, i.e. the deep bulk, the approximate thermofield double state will be dual to the two-sided planar AdS-Schwarzschild black brane \cite{Maldacena_2003} plus corrections due to the multicenter nature of the exterior solution. No known solution exists for the multicenter black brane geometry at nonzero temperature, so we solve for these corrections in perturbation theory. Matching these corrections in different coordinate patches glues together the wormhole solution. We will find that this gluing must introduce a global monodromy that inverts some spatial directions between the two throats in order to respect flux conservation. However, the full ten-dimensional spacetime remains globally orientable. The complete solution is unstable, as finite temperature breaks the supersymmetry of the BPS solution and turns on an effective potential for the scalar fields that break the $SU(2N)$ symmetry \cite{Kraus_1999, NAYEK2017192}. Supergravity wormhole solutions have been previously studied e.g. in \cite{BERGMAN2009300, Maldacena_2004, susywormhole}, but only in the case of connecting two different asymptotic spaces.

\begin{figure}[hbtp!]
\begin{center}
\includegraphics[width=.4\textwidth]{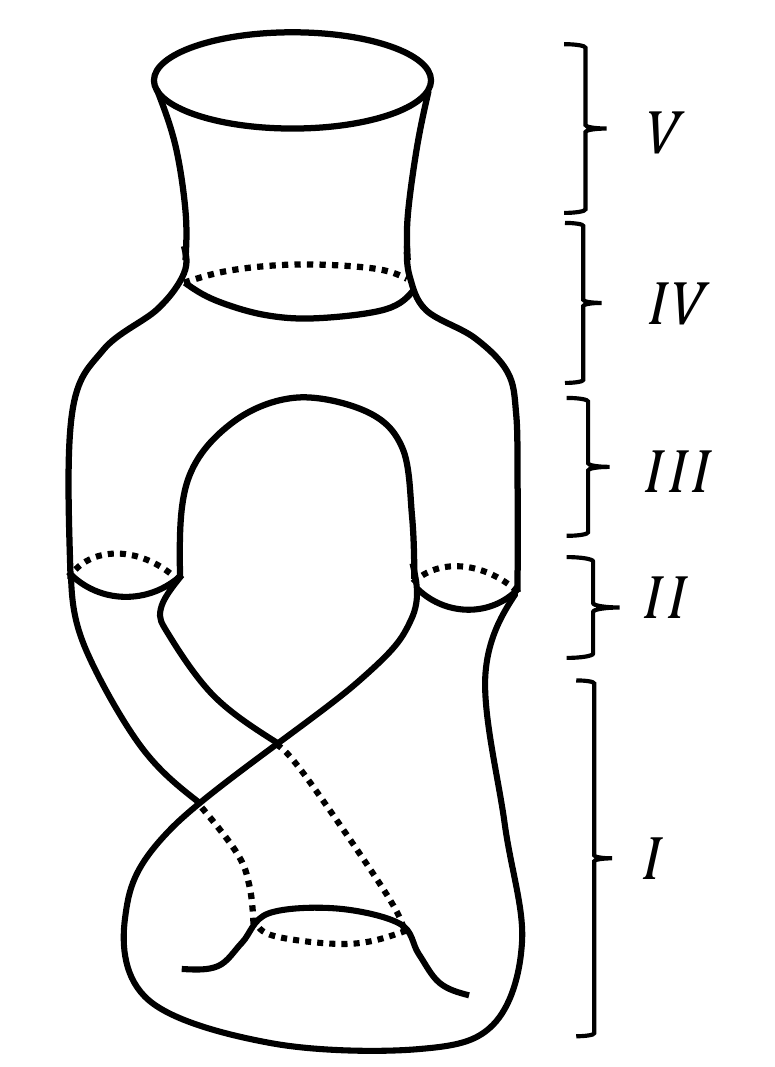}
\end{center}
\caption{Region I: Perturbatively corrected black brane glued with inversion to the rest of the geometry.  Region II: Linearized perturbations to vacuum AdS. Region III: Perturbatively corrected throat solution. Region IV: The (extremal) two-throat solution. Region V: Far from both throats, vacuum AdS with larger radius.\label{fig:wormhole2}}
\end{figure}

In Fig.~\ref{fig:wormhole2} we have labeled the different regimes in which a different coordinate patch or limit will be used to describe the solution. In region I, the solution is approximately the two-sided non-extremal black brane. Far from the horizon, in regions II and III, the effects of non-extremality are small and the solution is close to vacuum AdS$_5 \times S^5$. In this region, the perturbative corrections from the nonzero temperature and from the two throats can simultaneously be treated as linear corrections to the vacuum AdS$_5 \times S^5$ background, and therefore they linearly superpose. The leading corrections from the left throat are monopole corrections and do not break spherical symmetry around the right throat, which defines region II. However, we can include multipole effects from the left throat as linearized corrections, and these will be dominant over nonlinearities up to fourth order in the multipole expansion. This captures effects of the left throat breaking the spherical symmetry around the right throat, which defines region III. We will also be able to present solutions which are valid in regions I, II and parts of III simultaneously. These will be linearized perturbations of the finite temperature black brane geometry. These solutions show that leading multipole effects remain small near the causal horizons, but we find that they grow in the interior towards the singularity. Therefore the singularity in this wormhole is not of the AdS-Schwarzschild type. Finally, region IV is where both throats have non-perturbative effects, but their non-extremality is negligible and the solution is approximately the multicenter BPS solution, which in region V approaches that of pure AdS$_5 \times S^5$ with a larger AdS radius.

Entanglement between disconnected non-interacting boundary theories gives rise to wormholes where the boundaries are separated by causal horizons \cite{GALLOWAY2001255}. A large body of recent work has also been directed towards finding mechanisms that can create and send signals through traversable wormholes in the context of AdS/CFT. In general, supporting a traversable wormhole requires that one violate the averaged null energy condition (ANEC) \cite{Morris:1988tu, PhysRevLett.90.201102, PhysRevLett.61.1446, PhysRevLett.81.746, GALLOWAY2001255}, meaning that there exists an infinite null geodesic with tangent $k^{\mu}$ and affine parameter $\lambda$ such that \cite{GaoJafferisWall}\footnote{The authors of \cite{susywormhole} argued that this condition could be avoided for supersymmetric traversable wormholes connecting two asymptotic AdS universes in the context of pure gauged $\mathcal{N} = 2$ supergravity in four dimensions.}
\begin{align}
    \int_{-\infty}^{\infty} k^{\mu} k^{\nu} T_{\mu \nu} d\lambda < 0 \,.
\end{align}
Consequently, in order to build a traversable wormhole, there must be a negative source of stress-energy in the bulk. Several suggestions for introducing this negative stress-energy include inserting explicit double-trace couplings between the boundaries of the wormhole \cite{GaoJafferisWall, GJWBounds}, incorporating the perturbative gravitational back-reaction of bulk quantum fields \cite{Fu_2019, Fu_2019b}, including the Casimir energy of bulk fields running in non-contractible cycles \cite{SM_wormhole, selfsupp}, and nucleating and supporting wormholes via cosmic strings \cite{Horowitz_2019, Fu_2019a}. A particularly productive setting has been the correspondence between the SYK model and AdS$_2$ Jackiw-Teitelboim gravity \cite{doi:10.1002/prop.201700034, SYKwormhole, GaoJafferisSYK, SYKwormhole2}, which, while not an exact duality, has provided further support that explicit boundary couplings may render the bulk geometry traversable and provided an experimental setting by which probing wormhole traversability may be possible in the lab \cite{GJWprobes, wormhole_signaling}.

These constructions use the fact that the eternal AdS-Schwarzschild wormhole is marginally non-traversable in the sense that the null energy vanishes along the causal horizons, so arbitrarily small negative energy perturbations render the wormhole traversable. We will find that the leading classical corrections coming from the global structure of our single-boundary geometry preserve this marginal non-traversability.\footnote{We thank Simon Ross for discussions regarding this point.} This motivates us to describe a mechanism by which our single-boundary wormhole may become traversable by the presence of a natural ``double-trace" type operator in the IR of $\mathcal{N} = 4$ SYM generated by the Wilsonian RG flow \cite{connectivity}, although the presence of bulk fermions in the supergravity spectrum implies that other mechanisms mentioned above may also be a possibility depending, e.g., on the final sign of cancellations between Casimir energies.

The rest of this paper is organized as follows. In Sec.~\ref{sec:fieldtheory} we explain the pattern of symmetry breaking in the field theory and describe a particular entangled state in the IR. In Sec.~\ref{sec:wormholeGeomSoln} we solve for the metric and five-form of the wormhole solution in perturbation theory in type IIB supergravity and describe its global structure. In Sec.~\ref{sec:stability} we use the DBI action to estimate the instability timescale of the wormhole, and show that it is controlled by the same ratio of scales that governs the thermal effective potential in the SYM theory.  We also show that it is unlikely that the wormhole could be stabilized by adding rotation. In Sec.~\ref{sec:traverse} we discuss a mechanism for rendering our wormhole traversable, and in Sec.~\ref{sec:doublewormhole} we explain how to use our results to construct a double wormhole between two asymptotic universes. We conclude in the Discussion with comments and remarks for future directions of study.

\noindent\textbf{Conventions} \\
We work in ``mostly-plus" signature for Lorentzian metrics. The convention for five-form components is that
$F = \frac{1}{5!}F_{\mu \alpha \beta\gamma \delta} dx^{\mu} \wedge \ldots \wedge dx^{\delta} = F_{t123r} dt \wedge \ldots \wedge dr + F_{\theta_1 \ldots \theta_5} d\theta_1 \wedge \ldots \wedge  d\theta_5$ (all other components will be zero throughout this paper). The notation and combinatorial factors used in symmetrization of indices are for example $A_{(\mu}B_{\nu)} = \frac{1}{2!} (A_{\mu} B_{\nu} + A_{\nu} B_{\mu})$. The indices of all perturbative geometric quantities are raised with the background metric. In general this means index raising and lowering does not commute with perturbative variation. The action of the Hodge star on the components of $p$-forms in $d$ spacetime dimensions is $(\ast F)_{\nu_1 \ldots \nu_{d-p}} = \frac{1}{p!} \sqrt{-g} \epsilon_{\nu_1 \ldots \nu_{d-p} \sigma_1 \ldots \sigma_{p}} g^{\mu_1 \sigma_1} \ldots g^{\mu_p \sigma_p} F_{\mu_1 \ldots \mu_p}$ where $\epsilon_{\mu_1 \ldots \mu_n}$ is the Levi-Civita symbol and $\epsilon_{01\ldots (d-1)} = 1$.

\section{Description in Super-Yang Mills}\label{sec:fieldtheory}

The Lagrangian of $\mathcal{N} = 4$ SYM in terms of component fields is \cite{BRINK197777}
\begin{align}
    \mathcal{L}_0 &= \text{tr} \biggl(-\frac{1}{2g_{YM}^2} F_{\mu \nu} F^{\mu \nu} + \frac{\theta}{16\pi^2} F_{\mu \nu} \tilde{F}^{\mu \nu} - i\bar{\lambda}^a \bar{\sigma}^{\mu} D_{\mu} \lambda_a - \sum_i D_{\mu} \phi^i D^{\mu} \phi^i \nonumber \\
    &+g_{YM} \sum_{a,b,i} C^{ab}_i \lambda_a [\phi^i, \lambda_b] + g_{YM} \sum_{a,b,i} \bar{C}_{iab} \bar{\lambda}^a [\phi^i, \bar{\lambda}^b] + \frac{g_{YM}^2}{2} \sum_{i,j} [\phi^i, \phi^j]^2\biggr)\, ,
\end{align}
where $F_{\mu \nu} = \partial_{\mu} A_{\nu} - \partial_{\nu} A_{\mu} + i[A_{\mu}, A_{\nu}]$ is an $SU(2N)$ gauge field, $D_{\mu} \cdot = \partial_{\mu} \cdot + i[A_{\mu}, \cdot]$ is the covariant derivative on fields in the adjoint representation, $\lambda_a$ are four adjoint Weyl fermions, $\phi^i$ are six adjoint real scalars, and the $C^{ab}_i$ are the Clebsch-Gordon coefficients that couple two $\mathbf{4}$ representations of the $SU(4)_R$ symmetry to the $\mathbf{6}$ antisymmetric representation. The diagonal elements of the vacuum expectation value (vev) of the adjoint scalars $\phi^i$ in the AdS/CFT correspondence map to the positions of D3-branes in ten-dimensional flat space, while the off-diagonal elements are excitations of open strings stretching between the branes \cite{DouglasTaylor}. 

At zero temperature, any diagonal configuration of $\phi^i$ gives rise to a vanishing commutator in the potential for the scalars, so there is a large moduli space of stable vacua. In the dual gravity theory, this is equivalent to the statement that an arbitrary number of D3-branes can be superposed at any location in space without any force between them. We choose a vev that will correspond simply to separating two stacks of $N$ D3-branes in a single transverse coordinate by distance $\Lambda$, by expanding 
\begin{equation}
\phi^1 \to \psi + \varphi^1
\end{equation}
with the background $\psi$ given by
\begin{align}
    \psi = \frac{1}{2\pi \alpha'}\text{diag}(\Lambda \sqrt{N} , \Lambda \sqrt{N} , \ldots, 0, 0, \ldots) \, , \label{eq:scalarvev}
\end{align}
where the eigenvalue $\Lambda \sqrt{N} $ is repeated $N$ times and the vevs of all other $\phi^i$ are zero. 

We introduce an explicit factor of $\sqrt{N}$ to appropriately normalize the gauge-invariant classical observable $\text{tr} \, \psi^2$. In detail, we would like the classical value of the observable $\text{tr}\, \phi_1^2$, after proper normalization, to have a finite $\mathcal{O}(1)$ expectation in the large $N$ limit so that there is a well-defined classical gravitational dual.  Next we observe that $\text{tr}\, \varphi_1^2$ scales as $\mathcal{O}(N)$, since the components of $\varphi_1$ are $\mathcal{O}(1)$, so its connected two-point function scales as $\langle \text{tr}\, \varphi_1^2 \text{tr}\, \varphi_1^2 \rangle \sim \mathcal{O}(N^2)$.   But we know from large-N index counting that if the expectation value of a classical observable is taken to be $O(1)$, then the connected component of the two-point function of its quantum fluctuations should be $O(1/N^2)$.  Therefore, the observable $\text{tr}\, \phi_1^2$ requires an overall normalization proportional to $N^{-2}$. Consequently, to make the properly normalized value $N^{-2} \text{tr}\, \psi^2$ of the classical observable $\mathcal{O}(1)$ in the large $N$ limit, a factor of $\sqrt{N}$ should be included in $\psi$.

The factor of $\alpha' = \ell_s^2$ is required by dimensional analysis since the field $\phi^1$ has mass dimension one in four spacetime dimensions. This should be understood as a scale coming from open string theory, since it provides the energy cutoff such that the massless excitations of the open string endpoints moving on the D-brane world-volumes are described by $\mathcal{N}=4$ SYM\footnote{The $\mathcal{N}=4$ SYM does not include $\alpha'$ as a parameter. We introduce $\alpha'$ in order to set the dimensions of $\psi$ correctly as a fiducial scale where we expect the SYM description to break down, anticipating the correspondence with the supergravity description to be discussed later.}. The effective Lagrangian for fluctuations about this background is, rescaling the gauge field $A_{\mu} \to g_{YM} A_{\mu}$ to canonically normalize its kinetic term,
\begin{align}
    \mathcal{L} = \mathcal{L}_0\, +  \,\text{tr} \biggl(&2 i g_{YM} [\psi, A_{\mu}] \partial^{\mu} \varphi^1 + g_{YM}^2 ([\psi, A_{\mu}]^2 + 2 [\psi, A_{\mu}][\varphi^1, A^{\mu}]) +g_{YM} \sum_{a,b} C_1^{ab} \lambda_a [\psi, \lambda_b] \nonumber\\
    &+ g_{YM} \sum_{a,b} \bar{C}_{1ab} \bar{\lambda}^a [\psi, \bar{\lambda}^b]+g^2_{YM} \sum_i ([\psi, \phi^i]^2 + 2 [\psi, \phi^i][\varphi^1, \phi^i])\biggr)\, ,
\end{align}
where $\psi$ should be treated as a classical source. 

To understand this effective Lagrangian, it is instructive to expand the commutator of the vev $\psi$ with an arbitrary Hermitian matrix $M$ in the adjoint of $SU(2N)$, which may be written in block form as $M = \begin{pmatrix} M_A & M_B \\ M_B^{\dagger} & M_C \end{pmatrix}$ where each block is $N \times N$:
\begin{align}
    [\psi, M] &= \frac{\Lambda \sqrt{N} }{2\pi \alpha^{\prime} }\left(\begin{array}{cc} 0 & \:\:M_B \\  -M_B^{\dagger}  & \:\:0\end{array}\right) = \frac{\Lambda \sqrt{N}}{2\pi \alpha^{\prime} } M_{AO} \, .
\end{align}
We have labeled the final matrix $M_{AO}$ for the ``antihermiticized off-diagonal" piece of $M$. Note that $\text{tr}\, M_{AO}^2$ is strictly negative, which is required to give the correct signs below. Armed with this knowledge we further rewrite the commutators in the effective Lagrangian to make the dependence on the coupling $\Lambda$ clear, defining the 't Hooft coupling $\lambda = g_{YM}^2 N$ \cite{HOOFT1974461}:
\begin{align}
    \mathcal{L} = \mathcal{L}_0 \, + \, & \frac{1}{2\pi \alpha^{\prime}}\text{tr}\biggl(  \Lambda\sqrt{\lambda} \, A_{\mu, AO} \partial^{\mu} \varphi^1 + \frac{\lambda\Lambda^2 }{2\pi \alpha' } \,A_{\mu, AO}^2+ 2 \frac{\lambda \Lambda}{\sqrt{N}} \,A_{\mu, AO} [\varphi^1, A^{\mu}]+ \Lambda \sqrt{\lambda}\sum_{a,b} C_1^{ab}\, \lambda_a  \lambda_{b,AO} \nonumber \\
    &+ \Lambda \sqrt{\lambda} \sum_{a,b} \bar{C}_{1ab}\, \bar{\lambda}^a  \bar{\lambda}^b_{AO} + \frac{\lambda\Lambda^2 }{2\pi \alpha' } \sum_i  (\phi^{i}_{AO})^2 + 2 \frac{\lambda \Lambda}{\sqrt{N}} \sum_i  \phi^i_{AO}[\varphi^1, \phi^i] \biggr) \, . \label{eq:efflagrang}
\end{align}

The off-diagonal pieces of the gauge field, scalars, and fermions have acquired a mass \\ $ \Lambda  \sqrt{\lambda} / (2\pi \alpha^{\prime} ) =  \Lambda \sqrt{\lambda} / (2\pi \alpha^{\prime}) $.\footnote{This tree-level mass defines the Higgs scale at weak coupling. At strong coupling, the dependence on $\lambda$ may be different and we comment on this at the end of this section.} The diagonal blocks of the adjoint fields remain massless, so the background $\psi$ has Higgsed the theory $SU(2N) \to S(U(N) \times U(N))$. The extra terms remaining in \eqref{eq:efflagrang} coupling the gauge field to the scalar $\varphi^1$ are typical of those that appear in spontaneously broken non-Abelian gauge theories; we expect that there is a gauge choice which is an analog of the unitary gauge \cite{PhysRevLett.27.1688} where these terms vanish. 

In the symmetry-broken phase that we have chosen, low-lying excitations above the vacuum are local to only one of the $SU(N)$ factors of the gauge group. This is because, as we have shown above, the off-diagonal degrees of freedom can be made very heavy by choosing a large Higgs scale $\Lambda\sqrt{\lambda} / (2\pi \alpha')$. Therefore, at energies below the Higgs scale, the Hilbert space of the theory approximately factorizes into that of two separate $SU(N)$ gauge theories, each of which is dual to an AdS throat in the 10D supergravity. Following the ER=EPR conjecture, a state in SYM that possesses the appropriate entanglement between the $\mathcal{O} (N^2)$ light degrees of freedom in each $SU(N)$ factor should be dual to two AdS throats connected by a wormhole in the bulk \cite{doi:10.1002/prop.201300020}. Specifically, we build the approximate thermofield double state\footnote{In the UV it is not possible to factorize the Hilbert spaces due to the $SU(2N)$ being gauged, but it is approximately possible below the Higgs scale.} coupling the energy eigenstates of the Hamiltonian for the effective IR fields in each $SU(N)$ sector
\begin{align}
    |\text{TFD}_{\Lambda}\rangle = \sum_{i=1}^{E_i < E_c} e^{-\beta E_i / 2} |i\rangle_L \otimes |i^*\rangle_R \, ,\label{eq:tfdapprox}
\end{align}
where the sum runs over eigenstates of energy less than a cutoff energy scale $E_c$ set by the Higgs scale $\Lambda \sqrt{\lambda} / (2\pi \alpha')$, and the subscripts $L$ and $R$ refer to each of the two $SU(N)$ factors in the symmetry-broken theory. When the thermal energy density is much smaller than the cutoff energy density, we expect this state to be very close to the exact thermofield double state. In the deep bulk, this state is approximately dual to the planar two-sided AdS-Schwarzschild black brane which at fixed times describes a spacelike wormhole or ``Einstein-Rosen bridge" between two asymptotically AdS regions. However, in the ultraviolet of the field theory,  the state \eqref{eq:tfdapprox} is embedded in a single $SU(2N)$ SYM theory, so in fact the wormhole begins and ends in the same asymptotic region.

Let us be a bit more precise about how close \eqref{eq:tfdapprox} is to the thermofield double. The dominant contribution to \eqref{eq:tfdapprox} comes at energies where the Boltzman factor offsets the growth coming from the number of states. At low enough temperatures we can think of the individual $SU(N)$ theories as being conformal and therefore their canonical energy density and entropy are fixed by scale invariance and dimensional analysis\footnote{These relations are derived from $\log Z = c V\beta^{-3}$.}
\begin{equation}
\langle E \rangle_\beta/V = 3c\beta^{-4}, \quad S_{th}/V = 4c \beta^{-3}\, ,
\end{equation}
where $c$ is a constant proportional to the central charge and $V$ is the spatial volume. The contribution of a canonical window of states at the cutoff energy $E_c/V=3c \beta_c^{-4}$ is then estimated to be
\begin{equation}
\label{eq:corrections}
    e^{S_c-\beta E_c}\approx e^{V c \beta_c^{-4}(4 \beta_c-3 \beta)}\, ,
\end{equation}
i.e. we have an exponential suppression of these contributions if $\beta>4 \beta_c/3$. We take the cutoff temperature to be set by the Higgs scale $\beta_c^{-1} \lesssim \Lambda \sqrt{\lambda}/(2\pi \alpha')$. 

\begin{figure}[h!]
\begin{center}
\includegraphics[width=.5\textwidth]{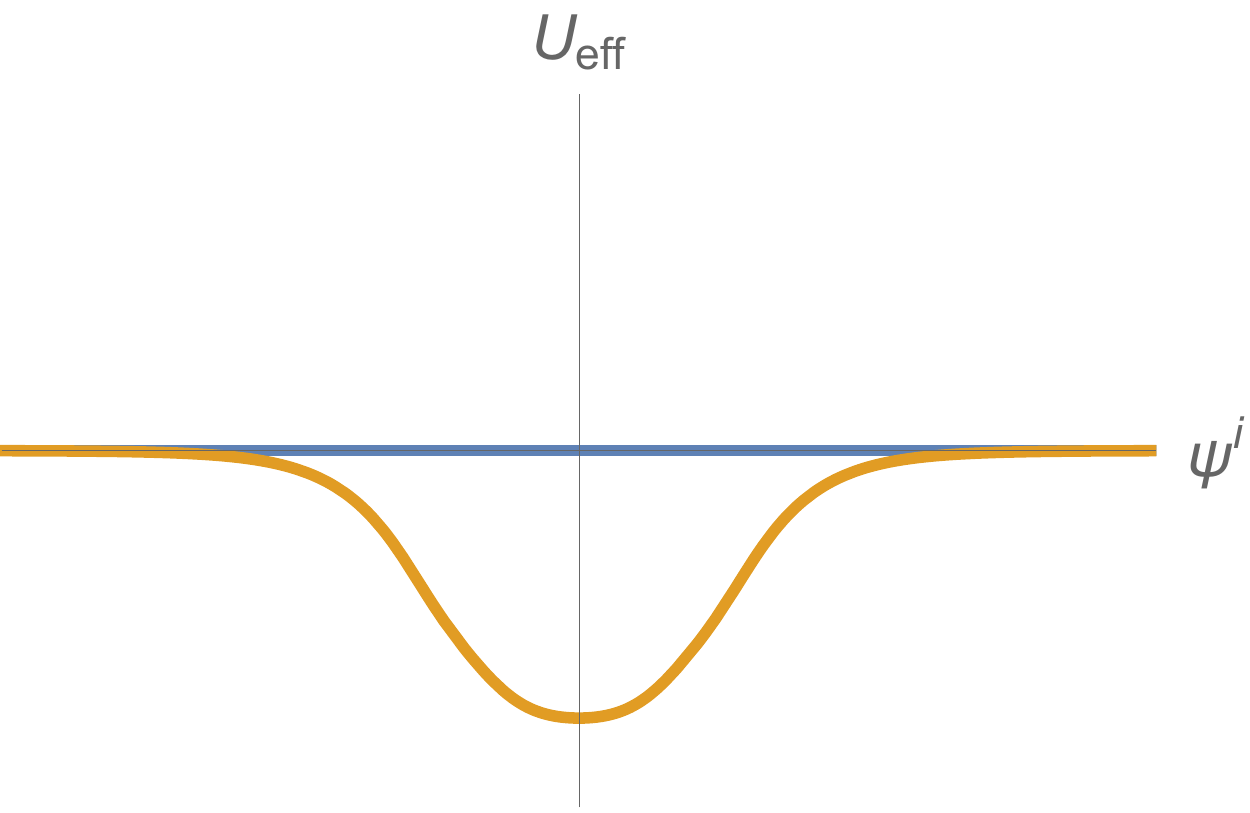}
\end{center}
\caption[The shape of the effective potential at finite temperature (orange) vs. zero temperature (blue). At zero temperature, the potential is flat and the components of the scalar background vev $\psi^i$ are free, while at finite temperature, the only stable configuration has all $\psi^i = 0$.]{The shape of the effective potential at finite temperature (orange) vs. zero temperature (blue). At zero temperature, the potential is flat and the components of the scalar background vev $\psi^i$ are free, while at finite temperature, the only stable configuration has all $\psi^i = 0$.\footnote[7]{\mbox{}}\label{fig:effpotential}}
\end{figure}

Reducing $|\text{TFD}_\Lambda\rangle$ onto the Hilbert space of either $SU(N)$ factor yields a state which looks approximately thermal in the infrared (up to corrections of order \eqref{eq:corrections}), with a temperature $\beta \sim r_0^{-1}$ that corresponds to a black brane of horizon radius $r_0$ in the gravity dual. In finite temperature field theory the supersymmetry of the SYM Lagrangian is broken and in particular the effective potential for the scalars $\phi^i$ is modified so that the only stable vacuum configuration is the one where the vevs of all the $\phi^i$ sit at the origin in moduli space. See Fig.~\ref{fig:effpotential} for a schematic depiction of the effective potential at finite temperature. Consequently, an initial configuration of the form \eqref{eq:scalarvev} is unstable and the vev will roll down the potential towards the origin. In the bulk dual, this has the well-known effect that nonextremal D-branes exert a nonzero attractive force on each other. Parametrically, at weak coupling $\lambda$ the thermal effective potential is controlled by the perturbatively small ratio of the thermal and Higgs scales, $\frac{2\pi \alpha'}{\beta \Lambda \sqrt{\lambda}}= \epsilon$.  The weak coupling effective potential cannot be directly compared with the dual semiclassical gravity, since the latter is only valid when the field theory coupling is strong. Nonetheless, in Sec.~\ref{sec:stability} we will estimate the timescale of the instability from the gravity dual using the DBI action of the underlying branes. In terms of bulk quantities in the gravity dual, the perturbative parameter controlling the thermal effective potential at small 't Hooft coupling is 
\begin{align}
    \epsilon =  \frac{\sqrt{2}}{\lambda} \frac{ r_0}{\Lambda }\, .
\end{align}
\setcounter{footnote}{7} \footnotetext{It may be surprising that the effective potential at zero temperature for the scalars is flat since the SYM theory is dual to the asymptotically AdS geometry that remains after the decoupling limit, and radially separated branes in asymptotically AdS space are subject to a potential barrier at infinity. However, the geometry sourced by two stacks of branes is a full ten-dimensional geometry that only approximately fibers into an AdS$_5$ and an $S^5$ close to each stack and near infinity. From the perspective of the asymptotic $S^5$, the branes are located at opposite poles and are not radially separated in the AdS space, so the AdS potential barrier does not apply.
} 
At \emph{strong} coupling in SYM, where the bulk dual admits a semiclassical description in supergravity, quantities computed at weak field theory coupling  are often rescaled by functions of $\lambda$ (see \cite{doi:10.1063/1.1372177} for a concrete example). Therefore, away from weak coupling we expect that the perturbative parameter that will control the bulk geometry in supergravity (and therefore parameterize the instability timescale of the wormhole) will take the form
\begin{align}
    \epsilon =  f(\lambda) \frac{\sqrt{2}}{\lambda} \frac{ r_0}{\Lambda } \, ,
\end{align}
for some function $f(\lambda)$. In the limit of large coupling $\lambda$ we will see that $f(\lambda) \sim \lambda$, so that the perturbative description of the classical geometry is naturally controlled by $\epsilon \sim r_0 / \Lambda$, which is independent of the string scale. It would be interesting to see if the function $f(\lambda)$ can be determined as an exact function of the coupling $\lambda$ using integrability techniques.

\section{Wormhole Geometry in Supergravity} \label{sec:wormholeGeomSoln}

In subsequent sections, we will write down the detailed solution to the equations of motion in each region. Our starting point for the construction, following \cite{pol95}, is the action of type IIB supergravity in string frame, restricted to the metric, dilaton, and five-form\footnote{The reviews \cite{DUFF1995213, peet} and textbooks \cite{kiritsis, johnson, amerd} provide compact and relevant introductions to D-brane solutions to type IIB supergravity that may be useful for subsequent sections.}:
\begin{align}
S_{\text{IIB}} = \frac{1}{2\kappa^2_{10}} \int d^{10} x \sqrt{-g} \left( e^{-2\phi} (R + 4\partial_{\mu} \phi \partial^{\mu} \phi) - \frac{1}{4\cdot 5!  } F_{\mu \nu \rho \sigma \tau} F^{\mu \nu \rho \sigma \tau}\right) \, ,\label{eq:IIBaction}
\end{align}
where $2\kappa_{10}^2 = (2\pi)^7 \alpha'^{4} g_s^2$. The asymptotic value of the dilaton has already been scaled out so that $e^{\phi} = 1$ at infinity. We work in the strongly coupled limit of the field theory, $g_s N \to \infty$, such that classical supergravity is valid. The background values of the various fermions of type IIB are taken to be zero self-consistently.

Taking as an ansatz that the dilaton will be constant everywhere so that we can drop terms involving its gradient, the classical equations of motion are
\begin{align}
e^{-2\phi} (R_{\mu \nu} - \frac12 R g_{\mu \nu}) &= -\frac{1}{8\cdot 5!} g_{\mu \nu} F_{\alpha \beta \gamma \delta \epsilon} F^{ \alpha \beta \gamma \delta \epsilon} + \frac{1}{4\cdot 4!} F_{\mu \alpha \beta \gamma \delta} F^{ \:\:\alpha \beta\gamma \delta}_{\nu} \label{eq:geomEOM1} \\
\partial_{\mu} (\sqrt{-g} F^{ \mu \nu \rho \sigma\tau}) &= 0 \, , \label{eq:maxwellbgd}
\end{align}
to be supplemented by the self-duality constraint $F = \ast F$, and with $e^{\phi}= 1$ everywhere. Due to the self-duality constraint, $ F_{\alpha \beta \gamma \delta \epsilon} F^{ \alpha \beta \gamma \delta \epsilon} \sim F \wedge \ast F = F \wedge F = 0$ since the wedge product is antisymmetric on five-forms. A straightforward computation by taking traces and using this identity shows that $R = 0$. Consequently, \eqref{eq:geomEOM1} simplifies to
\begin{align}
R_{\mu \nu}  &= \frac{1}{4\cdot 4!} F_{\mu \alpha \beta \gamma \delta} F^{ \:\:\alpha \beta\gamma \delta}_{\nu} \, . \label{eq:geomEOM2}
\end{align}

In subsequent sections we will write down perturbative corrections to solutions to the background equations of motion. We add these perturbative corrections at first order to the metric and five-form, $g_{\mu \nu} \to \bar{g}_{\mu \nu} + h_{\mu \nu}$ and $F \to \bar{F} + \delta F$, where the bar indicates quantities at background order, i.e. that solve \eqref{eq:maxwellbgd} and \eqref{eq:geomEOM2}. The perturbative equations of motion are
\begin{align}
    \nabla_{\lambda} \nabla_{(\mu} h^{\lambda}_{\:\:\nu)} - \frac12 \nabla_{\mu}\partial_{\nu} h - \frac12 \nabla_{\lambda} \nabla^{\lambda} h_{\mu \nu} &= \frac{1}{4 \cdot 4!} (\delta F_{\mu \alpha \beta \gamma \delta} \bar{F}_{\nu}^{\:\:\alpha \beta \gamma \delta} + \bar{F}_{\mu \alpha \beta \gamma \delta} \delta F_{\nu}^{\:\: \alpha \beta \gamma \delta}) \label{eq:perturbativeEOMs1}\\
    \partial_{\mu} \bigl[\sqrt{-\bar{g}} \bigl(\frac{h}{2} \bar{F}^{\mu \nu \rho \sigma \tau} + \delta F^{\mu \nu \rho \sigma \tau} \bigr) \bigr]&= 0 \, ,
    \label{eq:perturbativeEOMs2}
\end{align}
where $h = \bar{g}^{\mu \nu} h_{\mu \nu}$ is the trace of the metric perturbation and $\nabla_{\mu}$ is the covariant derivative with respect to $\bar{g}$. These equations must be supplemented with the self-duality constraint at all orders, such that $\bar{F} + \delta F= \ast (\bar{F} + \delta F)$, a nontrivial constraint since the Hodge dual involves the metric perturbations. The derivation of \eqref{eq:perturbativeEOMs1} and \eqref{eq:perturbativeEOMs2} can be found in Appendix~\ref{sec:perturbEOM}. In the following sections, we will exhibit solutions to the background equations of motion \eqref{eq:maxwellbgd} and \eqref{eq:geomEOM2} and to their first-order variation in \eqref{eq:perturbativeEOMs1} and \eqref{eq:perturbativeEOMs2}. Although we have labeled the regions of the geometry I - V in order of the flow from the IR to the UV in the field theory, we will describe the solutions below in a different order that will be more convenient for intuition.

\subsection{Region IV: Two-Center Harmonic Solution} \label{sec:twocenter}

The general two-center solution at nonzero temperature is not known even in perturbation theory, so we will first write down the background solution to \eqref{eq:maxwellbgd} and \eqref{eq:geomEOM2} without perturbations in this region. We begin with the solution corresponding to two stacks of $N$ extremal D3-branes placed at a separation $\Lambda$ in 10D Minkowski spacetime. The coordinates $t, x^1, \ldots, x^3$ extend parallel to the brane worldvolumes; we label the other six directions transverse to the branes as $r^1$ through $r^6$. Without loss of generality let the two stacks of branes be displaced in the $r^1$ direction. The solution in asymptotically flat space is BPS and the metric and five-form are given by \cite{DUFF1991409, HOROWITZ1991197}
\begin{align}
\label{eq:harmonicmulticenter}
ds^2 &= H^{-1/2} (- dt^2 + d\vec{x}^2) + H^{1/2} \delta_{ij} dr^i dr^j \\
F&=  (1+\ast) dt \wedge dx^1 \wedge dx^2 \wedge dx^3 \wedge dH^{-1} \, ,
\end{align}
where $H$ is a two-center harmonic function:
\begin{align}
H= 1 + \frac{L^4}{r^4} + \frac{L^4}{|\vec{r} \pm\vec{\Lambda}|^4} \, .
\end{align}
Here $\vec{\Lambda} = (\Lambda,0,0,0,0,0)$ and $r^2 = \sum_i (r^i)^2$. The choice of sign fixes the direction of displacement of the stacks of branes. 

The flux through the five-sphere jumps discontinuously when the radius of the five-sphere around one stack of branes crosses through the other stack. When the radius is smaller than $\Lambda$, the charge of a single stack is, by Stokes' theorem,
\begin{align}
    Q = \frac{1}{2\kappa^2}\int_{S^5} \ast F = \frac{L^4}{2g_s^2 (2\pi)^4 (\alpha')^4} \, .
\end{align}
The normalization comes from the normalization of the kinetic term for the five-form in \eqref{eq:IIBaction} \cite{pol95}. By the BPS condition, the charge is equal to the number of branes $N$ times the tension of a single extremal brane, $\tau =(2\pi)^{-3} (\alpha')^{-2} g_s^{-1}$, so
\begin{align*}
Q = N\tau \implies L^4 = 4\pi g_s N(\alpha')^2 \, .
\end{align*}
We now take the decoupling limit $\alpha' \to 0$ keeping fixed $\Lambda / \alpha'$ and $r^i / \alpha'$ \cite{costa}. Fixing $\Lambda/\alpha'$ amounts to fixing the Higgs scale of \eqref{eq:scalarvev} in the dual theory. To write a non-singular metric, we rescale $L^2 \to \alpha' L^2$, $r^i \to \alpha' r^i$, and $\Lambda \to \alpha' \Lambda$. The harmonic function becomes
\begin{align}
    H = \alpha'^{-2} \left(\frac{L^4}{r^4} + \frac{L^4}{|\vec{r} \pm\vec{\Lambda}|^4} \right) \, .
\end{align}
Lastly, we nondimensionalize coordinates by the rescaling
\begin{align}
    \frac{r_0 t}{L^2} = \tilde{t}, \qquad \frac{r}{r_0} = \tilde{r}, \qquad \frac{r_0 x^i}{L^2} = \tilde{x}^i \, . \label{eq:nondim}
\end{align}
Here $r_0$ is the wormhole horizon radius, to be introduced in subsequent sections. This nondimensionalization will be convenient in other regions where it removes the length scale $r_0$. We introduce the parameter $\epsilon = \frac{r_0}{ \Lambda}$. The full wormhole solution will only be valid in the limit $\epsilon \ll 1$ where the horizons of the two stacks of branes are well-separated. Given these definitions, the full solution in region IV is
\begin{align}
\frac{1}{\alpha'}ds^2 &= L^2 \biggl[\biggl(\frac{1}{\tilde{r}^4} + \frac{1}{|\vec{\tilde{r}} \pm \vec{\epsilon}^{-1}|^4}  \biggr)^{-1/2} (- d\tilde{t}^2 + d\vec{\tilde{x}}^2) +  \biggl(\frac{1}{\tilde{r}^4} + \frac{1}{|\vec{\tilde{r}} \pm \vec{\epsilon}^{-1}|^4}  \biggr)^{1/2} \delta_{ij} d\tilde{r}^i d\tilde{r}^j\biggr] \\
\frac{1}{\alpha'^2} F&= L^4 \biggl[ (1+\ast) d\tilde{t} \wedge d\tilde{x}^1 \wedge d\tilde{x}^2 \wedge d\tilde{x}^3 \wedge d \biggl(\frac{1}{\tilde{r}^4} + \frac{1}{|\vec{\tilde{r}} \pm \vec{\epsilon}^{-1}|^4}  \biggr)^{-1}\biggr] \, ,
\end{align}
where $\vec{\epsilon}^{-1} = (\epsilon^{-1},0,0,0,0,0)$.

It is convenient for subsequent sections to expand this solution close to the stack of branes at the origin, $r^i \ll \Lambda$. In this limit the (nondimensionalized) harmonic function becomes simply
\begin{align}
    H = \frac{1}{\tilde{r}^4} + \epsilon^4 \, .\label{eq:Hlimit}
\end{align}
That is, all dependence on $r^1$ is subleading, so spherical symmetry about the stack of branes is valid in this limit. Furthermore, this expansion is valid around either stack provided the radial coordinate is defined appropriately. Defining the hyperspherical coordinates
\begin{align}
\begin{aligned}
r_1 &= r\cos \theta_1 \\
r_2 &=r\cos \theta_2\sin \theta_1 \\
r_3 &= r\cos \theta_3\sin \theta_2\sin \theta_1\\
r_4 &= r\cos \theta_4\sin \theta_3\sin \theta_2\sin \theta_1\\
r_5 &= r\cos \theta_5\sin \theta_4\sin \theta_3\sin \theta_2\sin \theta_1\\
r_6 &= r\sin \theta_5\sin \theta_4\sin \theta_3\sin \theta_2\sin \theta_1 \, ,
\end{aligned}
\end{align}
and series expanding $H^{\pm 1/2}$ yields
\begin{align}
\frac{1}{\alpha'}  ds^2
&= L^2\biggl[ \tilde{r}^2 \left(1- \frac12 (\epsilon \tilde{r})^4\right)(-d\tilde{t}^2+d\vec{\tilde{x}}^2) + \left(1+\frac12 (\epsilon \tilde{r})^4 \right)(\frac{d\tilde{r}^2}{\tilde{r}^2} +  d\Omega_5^2) \biggr] \label{eq:throatcorrect1}\\
\frac{1}{\alpha'^2} F &=  4L^4 \biggl[\tilde{r}^3 \left(1 - 2(\epsilon \tilde{r})^4\right)  d\tilde{t} \wedge d\tilde{x}^1 \wedge d\tilde{x}^2 \wedge d\tilde{x}^3 \wedge d\tilde{r} \nonumber\\
&\qquad\qquad\qquad+ \sin^4 \theta_1 \sin^3 \theta_2 \sin^2 \theta_3 \sin \theta_4 d\theta_1 \wedge d\theta_2 \wedge d\theta_3 \wedge d\theta_4 \wedge d\theta_5\biggr] \, .\label{eq:throatcorrect2}
\end{align}
In this form, the linearized corrections to vacuum AdS$_5 \times S^5$ deep within a single throat are apparent. These corrections will be useful in subsequent sections.

\subsection{Region V: Asymptotics}

In the region far from both stacks of branes, we take the limit $r^i \gg \Lambda$, by which the harmonic function $H$ simplifies to
\begin{align}
    H = \frac{2L^4}{r^4}
\end{align}
Writing $L_{\infty}^4 = 2L^4$, taking the decoupling limit, rescaling, and nondimensionalizing, the solution in region V is
\begin{align}
  \frac{1}{\alpha'}  ds^2 &=   L_{\infty}^2\biggl[ \tilde{r}^2 (-d\tilde{t}^2+d\vec{\tilde{x}}^2) +  \frac{d\tilde{r}^2}{\tilde{r}^2} +  d\Omega_5^2 \biggr] \\
\frac{1}{\alpha'^2} F &=  4L_{\infty}^4 \biggl[\tilde{r}^3  d\tilde{t} \wedge d\tilde{x}^1 \wedge d\tilde{x}^2 \wedge d\tilde{x}^3 \wedge d\tilde{r} + \sin^4 \theta_1 \sin^3 \theta_2 \sin^2 \theta_3 \sin \theta_4 d\theta_1 \wedge d\theta_2 \wedge d\theta_3 \wedge d\theta_4 \wedge d\theta_5\biggr] \, .
\end{align}
This solution is vacuum AdS$_5 \times S^5$ with AdS length $L_{\infty}^4 = 2L^4$. The charge is
\begin{align}
Q  = \frac{L_{\infty}^4}{2g_s^2 (2\pi)^4 (\alpha')^4} \, ,
\end{align}
leading to
\begin{align}
    L_{\infty}^4 = 4\pi g_s (2N)(\alpha')^2  
\end{align}
from the flux quantization condition. At infinity, the flux sees both stacks of branes as if they are at the origin, as expected. This region corresponds to the UV in the field theory where the $SU(2N)$ symmetry is unbroken.

\subsection{Region I: Black Brane}

In region I, the solution is the geometry of the two-sided black brane with perturbative corrections coming from the second throat in the full geometry. In this region, the solution will describe the geometry close to one of the two stacks with the origin of coordinates placed at the location of the stack, that is, in the limit $r^i \ll \Lambda$ of Sec.~\ref{sec:twocenter}. The solution preserves the $SO(3,1) \times SO(6)$ isometries induced by the brane locations. Following the conventions of \cite{peet}, the metric and five-form of the asymptotically flat solution are \cite{HOROWITZ1991197, DUFF1991409, DUFF1995213}
\begin{align}
ds^2 &= H(r)^{-1/2} (-f(r) dt^2 + d\vec{x}^2) + H(r)^{1/2} (dr^2 / f(r) + r^2 d\Omega_5^2) \\
F&= \sqrt{1+\frac{r_0^4}{L^4}}(1+\ast) dt \wedge dx^1 \wedge dx^2 \wedge dx^3 \wedge dH^{-1} \, ,
\label{eq:flatnonextremal}
\end{align}
with $H(r) = 1+\frac{L^4}{r^4}$ and $f(r) = 1- \frac{r_0^4}{r^4}$. As the horizon radius $r_0 \to 0$ one approaches the extremal limit of the brane solution. The five-form can be written explicitly in coordinates as
\begin{align}
F&= \sqrt{1+\frac{r_0^4}{L^4}} \biggl[\frac{4L^4}{ r^5 H(r)^2} dt \wedge dx^1 \wedge dx^2 \wedge dx^3 \wedge dr \nonumber \\
& \qquad\qquad\qquad\qquad + 4L^4 \sin^4 \theta_1 \sin^3 \theta_2 \sin^2 \theta_3 \sin \theta_4 d\theta_1 \wedge d\theta_2 \wedge d\theta_3 \wedge d\theta_4 \wedge d\theta_5\biggr] \, .
\end{align}
Consequently, the charge is
\begin{align}
Q = \frac{1}{2\kappa^2} \int_{S^5} \ast F = \frac{L^4}{2g_s^2 (2\pi)^4 (\alpha')^4 }\sqrt{1+\frac{r_0^4}{L^4}} \, .
\end{align}
The charge remains equal to the number of branes $N$ times the tension $\tau$ of an extremal brane, so
\begin{align}
Q = N\tau \implies L^4 = -\frac12 r_0^4 + \sqrt{(4\pi g_s N(\alpha')^2 )^2 + \frac14 r_0^8} \, .
\end{align}

We now take the decoupling limit $\alpha' \to 0$ keeping fixed $r^i / \alpha'$ and $r_0 / \alpha'$, rescaling $L^2 \to \alpha' L^2$, $r_0\rightarrow \alpha' r_0$ and $r \to \alpha' r $. The resulting solution has the same form as \eqref{eq:flatnonextremal} with $H=L^4/r^4$, no overall scaling on the five-form, and $f(r)$ unchanged. Nondimensionalizing following \eqref{eq:nondim} it can be written as
\begin{align}
\frac{1}{\alpha'} ds^2 &= L^2 \biggl[ -\tilde{r}^2(1-\frac{1}{\tilde{r}^4}) d\tilde{t}^2 +  \tilde{r}^2 d\vec{\tilde{x}}^2 + \frac{d\tilde{r}^2}{\tilde{r}^2(1-\frac{1}{\tilde{r}^4})} +  d\Omega_5^2 \biggr] \label{eq:blackbranebgdgeom} \\
\frac{1}{\alpha'^{2}} F &= 4L^4 \biggl[\tilde{r}^3 d\tilde{t} \wedge d\tilde{x}^1 \wedge d\tilde{x}^2 \wedge d\tilde{x}^3 \wedge d\tilde{r} + \sin^4 \theta_1 \sin^3 \theta_2 \sin^2 \theta_3 \sin \theta_4 d\theta_1 \wedge d\theta_2 \wedge d\theta_3 \wedge d\theta_4 \wedge d\theta_5 \biggr] \, . \label{eq:blackbranebgdform}
\end{align}
This is the background solution in region I. Note that the finite temperature factor has dropped out of the five-form after the decoupling limit. Consequently, the charge in the decoupling limit is simply
\begin{align}
    Q = \frac{L^4}{2g_s^2 (2\pi)^4 (\alpha')^4 } \, .
\end{align}
This leads to the extremal quantization condition for the black brane in AdS,
\begin{align}
L^4 =  4\pi g_s N(\alpha')^2  \, .
\end{align}
In Sec.~\ref{sec:monopole} we will describe the perturbative corrections to \eqref{eq:blackbranebgdgeom}, \eqref{eq:blackbranebgdform} coming from the second throat, though we first describe the general structure of the multipole expansion that gives rise to these perturbative corrections in Sec.~\ref{sec:linearized}.

\subsection{Regions II-III: Linearized Regime} \label{sec:linearized}

Regions II and III are the intermediate regimes far from the horizon and sufficiently deep within a single throat such that the corrections to vacuum AdS$_5 \times S^5$ both from the throat and from the wormhole can be linearized. These regions are defined by $\tilde{r} \sim \mathcal{O}(\epsilon^{-1/2})$, where the background is empty AdS and corrections to this coming both from the harmonic function and the blackening factor are $\mathcal{O}(\epsilon^2)$. Since both of these corrections can be treated as linearized and the equations of motion are linear in the perturbations, the full solution can be written simply as the linear superposition of the two,
\begin{align}
\frac{1}{\alpha'}ds^2 &=H_0(r)^{-1/2} \left[ 1-\frac{1}{2}\frac{\delta H(r,\theta_1)}{H_0(r)}\right] \left(-\left[1-\frac{r_0^4}{r^4} \right] dt^2 + d\vec{x}^2 \right) \nonumber \\ &+ H_0(r)^{1/2} \left[ 1+\frac{1}{2}\frac{\delta H(r,\theta_1)}{H_0(r)}\right] \left(dr^2 \left[1+\frac{r_0^4}{r^4} \right] + r^2 d\Omega_5^2\right) \\
\frac{1}{{\alpha'}^2}F&=  (1+\ast) dt \wedge dx^1 \wedge dx^2 \wedge dx^3 \wedge d\left[ H_0(r)^{-1}\left( 1-\frac{\delta H(r,\theta_1)}{H_0(r)}\right)\right] \, ,
\end{align}
where $H_0(r)=\frac{L^4}{r^4}$ and
\begin{equation}
\label{eq:harmonicmultipole}
    \delta H(r,\theta_1)=\frac{L^4}{\Lambda ^4}+\frac{4 L^4 r \cos (\theta_1 )}{\Lambda ^5} +\frac{2 L^4 r^2 (3 \cos (2 \theta_1 )+2)}{\Lambda ^6}+\frac{4 L^4 r^3 (3 \cos (\theta_1 )+2 \cos (3 \theta_1 ))}{\Lambda ^7}+ \mathcal{O}(\Lambda^{-8}) \, ,
\end{equation}
and the Hodge star must be applied so that in the result we linearize both in $\delta H$ and $r_0^4$. This solves the linearized equations of motions simply because it is the sum of two linear perturbations of AdS$_5\times S^5$, one defined by expanding \eqref{eq:harmonicmulticenter} in $\delta H$ with $H=H_0+\delta H$, and the other by expanding \eqref{eq:blackbranebgdgeom} in $r_0^4/r^4$. Note that the leading nonlinearity from the two centered harmonic function comes at $(\delta H)^2 \sim \Lambda^{-8}$ and therefore we can keep the multipole expansion \eqref{eq:harmonicmultipole} up to $\mathcal{O}(\Lambda^{-7})$ in the linearized regime.\footnote{After nondimensionalizing, the multipole expansion is controlled by powers of $\epsilon$, so the leading nonlinearity is at $\mathcal{O}(\epsilon^8)$.}

We define region II as the patch where spherical symmetry around the throat is approximately unbroken and hence we can stop in the multipole expansion of $\delta H (r, \theta_1)$ at monopole order. This means keeping only the $L^4/\Lambda^4$ term in \eqref{eq:harmonicmultipole}. In this case the solution explicitly reads in nondimensionalized coordinates
\begin{align}
\frac{1}{\alpha'} ds^2 &= L^2 \biggl[-\tilde{r}^2\left(1- \frac{(\epsilon \tilde{r})^4}{2}  -\frac{1}{\tilde{r}^4} \right) d\tilde{t}^2 +\tilde{r}^2\left(1- \frac{(\epsilon \tilde{r})^4}{2}\right) d\vec{\tilde{x}}^2 \nonumber\\ &\qquad\qquad\qquad\qquad\qquad\qquad\qquad  + \frac{1}{\tilde{r}^2}\left(1+\frac{(\epsilon \tilde{r})^4}{2} + \frac{1}{\tilde{r}^4}\right)d\tilde{r}^2 +  \left(1+\frac{(\epsilon \tilde{r})^4}{2} \right) d\Omega_5^2 \biggr] \label{eq:linear1}\\
\frac{1}{\alpha^{'2}} F &= 4L^4 \biggl[\tilde{r}^3 \left(1 - 2(\epsilon \tilde{r})^4 \right)  d\tilde{t} \wedge d\tilde{x}^1 \wedge d\tilde{x}^2 \wedge d\tilde{x}^3 \wedge d\tilde{r} \nonumber  \\&\qquad\qquad\qquad\qquad\qquad+  \sin^4 \theta_1 \sin^3 \theta_2 \sin^2 \theta_3 \sin \theta_4 d\theta_1 \wedge d\theta_2 \wedge d\theta_3 \wedge d\theta_4 \wedge d\theta_5 \biggr] \, .\label{eq:linear2}
\end{align}
In this regime, the equations of motion can be solved by hand; see Appendix~\ref{sec:linearizedApp}. The procedure involves several undetermined constants and an undetermined function as a consequence of a residual diffeomorphism freedom.

\subsection{Joint solution in regions I-II: Monopole contribution}
\label{sec:monopole}

Now, we solve \eqref{eq:perturbativeEOMs1} and \eqref{eq:perturbativeEOMs2} for the perturbations to the metric and the five-form in the backgrounds of \eqref{eq:blackbranebgdgeom} and \eqref{eq:blackbranebgdform}, i.e. the non-extremal black brane. In this subsection we deal with the case when spherical symmetry is intact, that is, we solve for the monopole contribution of the far throat down the near throat. We begin with an ansatz for the perturbations consistent with the $SO(3,1) \times SO(6)$ symmetry
\begin{align}
    \frac{1}{\alpha'}ds^2 &= L^2 \left[-\tilde{r}^2(1-\frac{1}{\tilde{r}^4})(1+\delta g_{\tilde{t}\tilde{t}}) d\tilde{t}^2 + \tilde{r}^2 (1+ \delta g_{\tilde{i}\tilde{i}}) d\vec{\tilde{x}}^2 + \frac{d\tilde{r}^2}{\tilde{r}^2(1-\frac{1}{\tilde{r}^4})} (1+\delta g_{\tilde{r}\tilde{r}}) + (1+\delta g_{\Omega \Omega}) d\Omega_5^2 \right] \\
    \frac{1}{\alpha^{'2}} F &=4L^4\biggl[\tilde{r}^3 (1+a(\tilde{r})) d\tilde{t} \wedge d\tilde{x}^1 \wedge d\tilde{x}^2 \wedge d\tilde{x}^3 \wedge d\tilde{r} \nonumber \\
&\qquad + (1+b(\tilde{r}))\sin^4 \theta_1 \sin^3 \theta_2 \sin^2 \theta_3 \sin \theta_4 d\theta_1 \wedge d\theta_2 \wedge d\theta_3 \wedge d\theta_4 \wedge d\theta_5\biggr] \, ,
\end{align}
where the metric perturbations are all functions only of the radial coordinate $r$. In terms of the perturbations, the Maxwell equations and self-duality constraint reduce simply to
\begin{align}
    2a-2b-3\delta g_{\tilde{i}\tilde{i}} - \delta g_{\tilde{t}\tilde{t}}-\delta g_{\tilde{r}\tilde{r}}+5\delta g_{\Omega \Omega} &= 0 \label{eq:selfdualBB} \\
    2a'-3\delta g_{\tilde{i}\tilde{i}}' - \delta g_{\tilde{t}\tilde{t}}'-\delta g_{\tilde{r}\tilde{r}}'+5\delta g_{\Omega \Omega}' &= 0 \, .\label{eq:maxwellBB}
\end{align}
The geometric equations of motion are:
\begin{align}
\begin{aligned}
    &\left(1-5 \tilde{r}^4\right) \delta g_{\Omega \Omega }'- \tilde{r} \left(\tilde{r}^4-1\right) \delta g_{\Omega \Omega }''-16\tilde{r}^3 (b-2 \delta g_{\Omega \Omega }) = 0  \\
  &-16\tilde{r}^3 a + 24 \tilde{r}^3 \delta g_{\tilde{i}\tilde{i}} + 8\tilde{r}^3 \delta g_{\tilde{t}\tilde{t}} + 3(\tilde{r}^4 + 1)\delta g_{\tilde{i}\tilde{i}}' + 6\tilde{r}^4 \delta g_{\tilde{t}\tilde{t}}' \\
  &\qquad\qquad\qquad- (\tilde{r}^4 + 1)\delta g_{\tilde{r}\tilde{r}}' + 5(\tilde{r}^4 + 1) \delta g_{\Omega \Omega}' + (\tilde{r}^5 - \tilde{r} ) \delta g_{\tilde{t}\tilde{t}}'' = 0 \\
   &-16\tilde{r}^3 a + 24 \tilde{r}^3 \delta g_{\tilde{i}\tilde{i}} + 8\tilde{r}^3 \delta g_{\tilde{t}\tilde{t}} -4(2\tilde{r}^4-1) \delta g_{\tilde{i}\tilde{i}}'- (\tilde{r}^4-1)\delta g_{\tilde{t}\tilde{t}}' \\&\qquad\qquad\qquad+(\tilde{r}^4-1)\delta g_{\tilde{r}\tilde{r}}' - 5(\tilde{r}^4-1)\delta g_{\Omega \Omega}' - (\tilde{r}^5 - \tilde{r}) \delta g_{\tilde{i}\tilde{i}}'' = 0 \\
  &-16\tilde{r}^3 a + 24 \tilde{r}^3 \delta g_{\tilde{i}\tilde{i}} + 8\tilde{r}^3 \delta g_{\tilde{t}\tilde{t}} + 3 (3\tilde{r}^4-1)\delta g_{\tilde{i}\tilde{i}}' + 3(\tilde{r}^4+1)\delta g_{\tilde{t}\tilde{t}}'-2(2\tilde{r}^4-1) \delta g_{\tilde{r}\tilde{r}}' \\
  &\qquad\qquad\qquad+5(\tilde{r}^4+1)\delta g_{\Omega \Omega}' + 3(\tilde{r}^5-\tilde{r})\delta g_{\tilde{i}\tilde{i}}'' + (\tilde{r}^5-\tilde{r})\delta g_{\tilde{t}\tilde{t}}''+5(\tilde{r}^5-\tilde{r}) \delta g_{\Omega \Omega}'' = 0 \, .
  \end{aligned}
\end{align}
Note that we are in the region $\tilde{r}>1$ outside the horizon at all times. We solve the equations of motion by the ansatz that the perturbations will be equal to the near-horizon corrections of the two-throat solution given in \eqref{eq:throatcorrect1} and \eqref{eq:throatcorrect2}. This ansatz turns out to be correct provided that we include an additional contribution to the five-form perturbations $a(\tilde{r})$ and $b(\tilde{r})$. The full perturbative solution in region I is 
\begin{align}
\label{eq:monopole}
    \frac{1}{\alpha'}ds^2 &= L^2 \biggl[-\tilde{r}^2(1-\frac{1}{\tilde{r}^4})(1-\frac12 (\epsilon \tilde{r})^4) d\tilde{t}^2 + \tilde{r}^2 (1-\frac12 (\epsilon \tilde{r})^4 ) d\vec{\tilde{x}}^2 \nonumber\\
    &\qquad\qquad\qquad\qquad\qquad\qquad\qquad + \frac{d\tilde{r}^2}{\tilde{r}^2(1-\frac{1}{\tilde{r}^4})} (1+\frac12 (\epsilon \tilde{r})^4) + (1+\frac12 (\epsilon \tilde{r})^4 ) d\Omega_5^2 \biggr] \\
    \frac{1}{\alpha^{'2}} F &=4L^4\biggl[\tilde{r}^3 \left(1-2(\epsilon \tilde{r})^4 + \frac{\epsilon^4}{2}\right) d\tilde{t} \wedge d\tilde{x}^1 \wedge d\tilde{x}^2 \wedge d\tilde{x}^3 \wedge d\tilde{r}  \nonumber \\
&\qquad\qquad\qquad\qquad\qquad +(1+\frac{\epsilon^4}{2})\sin^4 \theta_1 \sin^3 \theta_2 \sin^2 \theta_3 \sin \theta_4 d\theta_1 \wedge d\theta_2 \wedge d\theta_3 \wedge d\theta_4 \wedge d\theta_5\biggr] \, .
\end{align}
In this regime, we are close to the horizon, so $\tilde{r} \sim \mathcal{O}(1)$ and the perturbative corrections are $\mathcal{O}(\epsilon^4)$. When $\tilde{r}$ gets large, this solution matches onto \eqref{eq:linear1} in the linearized regime, region II.

The physical interpretation of the leading monopole contribution from the presence of the other throat is to create a small region of flat space around the black brane. This can be seen by noting that  \eqref{eq:monopole} can be obtained by linearizing in $1/\Lambda^4$ the non-perturbative solution (in dimensionful coordinates):
\begin{align}
\frac{1}{\alpha'}ds^2 &=\left(\frac{L^4}{r^4}+\frac{L^4}{\Lambda^4} \right)^{-1/2}  \left(-\left[1-\frac{r_0^4}{r^4} \right] dt^2 + d\vec{x}^2 \right) + \left(\frac{L^4}{r^4}+\frac{L^4}{\Lambda^4} \right)^{1/2}  \left(dr^2 \left[1-\frac{r_0^4}{r^4} \right]^{-1} + r^2 d\Omega_5^2\right) \label{eq:metricnonpert}\\
\frac{1}{{\alpha'}^2}F&= \sqrt{1+\frac{r_0^4}{\Lambda^4}} (1+\ast) dt \wedge dx^1 \wedge dx^2 \wedge dx^3 \wedge d\left(\frac{L^4}{r^4}+\frac{L^4}{\Lambda^4} \right)^{-1} \label{eq:fiveformnonpert}\, ,
\end{align}
which is obtained by truncating the multipole expansion at monopole order but keeping the nonlinear dependence on the harmonic functions, as well as the blackening factor. It is easy to see that this solution is just a single non-extremal black brane in asymptotically flat space, in rescaled coordinates $t'=\frac{\Lambda}{L}t$, $\vec{x}'=\frac{\Lambda}{L}\vec{x}$, $r'=\frac{L}{\Lambda}r$. This rescaling puts the solution \eqref{eq:metricnonpert} in the form \eqref{eq:flatnonextremal} but with a rescaled horizon radius $r_0'= \frac{L}{\Lambda}r_0$.

\subsection{Joint solution in regions I-II-III: Dipole contribution}
\label{sec:dipole}

It is interesting to ask if we can capture the leading effect of spherical symmetry breaking on the wormhole. The above solutions contain the monopole contribution from the presence of the other throat. At next order, there is a dipole contribution from the harmonic function
\begin{equation}
    H=L^4\left(\frac{1}{r^4} + \frac{1}{\Lambda^4} + 4 \frac{ r \cos \theta_1}{\Lambda^5} + \cdots \right)\, , \label{eq:dipoleharmonic}
\end{equation}
We look for a solution including dipole effects in all three regions I-III, that is, we keep the blackening factor exact. We take a general ansatz where the harmonic functions $H_g$ in the metric and $H_F$ in the five-form are allowed to be different,
\begin{align}
\label{eq:dipole}
\frac{1}{\alpha'}ds^2 &= H_g(r,\theta_1)^{-1/2} (-f(r) dt^2 + d\vec{x}^2) + H_g(r,\theta_1)^{1/2} (dr^2 / f(r) + r^2 d\Omega_5^2) \\
\frac{1}{{\alpha'}^2}F&= B(1+\ast) dt \wedge dx^1 \wedge dx^2 \wedge dx^3 \wedge dH_F^{-1} \, ,
\end{align}
with $f(r)=1-r_0^4/r^4$ and $B=1+\frac{r_0^4}{2  \Lambda^4}$. Similarly, we take a general ansatz for $H_g$ and $H_F$ whereby both must be asymptotically equal to \eqref{eq:dipoleharmonic} as $r\to\infty$,
\begin{align}
    H_g &= L^4\left(\frac{1}{r^4} + \frac{1}{\Lambda^4} + 4 \frac{ h_g(r) \cos \theta_1}{\Lambda^5} + \cdots \right) \\
    H_F &= L^4\left(\frac{1}{r^4} + \frac{1}{\Lambda^4} + 4 \frac{ h_F(r) \cos \theta_1}{\Lambda^5} + \cdots \right)\, ,
\end{align}
that is, $h_g\sim h_F \sim r$ as $r\rightarrow \infty$. Requiring the $r\theta_1$ component of the curvature equation of motion \eqref{eq:geomEOM2} to vanish at order $1/\Lambda^5$ gives
\begin{equation}
    h_F(r)=\frac{2r^4-r_0^4}{2r^4}h_g(r) \, .
\end{equation}
Imposing this, it turns out all the remaining components of \eqref{eq:geomEOM2}, as well as the only non-vanishing component of Maxwell's equation, $(d F)_{r \theta_1...\theta_5}=0$ are proportional to the equation
\begin{equation}
    -5 r^3 h_g+(r^4-r_0^4)(5 h_g'+r h_g'')=0 \, .
\end{equation}
This is a second order equation with two initial conditions. One is fixed by $h_g(r\rightarrow \infty ) \rightarrow r$. The other is fixed by requiring the solution to stay real in the interior of the wormhole, $r<r_0$. It turns out that the latter condition translates into $h_g(r_0)=0$, so that the location of the horizon is not affected by the perturbation. The solution is then
\begin{align}
    h_g(r)&=r_0 Q\left( \frac{r_0^4}{r^4}\right), \\
    Q(x)&=\frac{\, _2F_1\left(-\frac{5}{4},-\frac{1}{4};-\frac{1}{2};x\right)+\frac{8 x^{3/2} \Gamma \left(\frac{5}{4}\right) \Gamma \left(\frac{9}{4}\right) }{3 \Gamma \left(-\frac{1}{4}\right) \Gamma \left(\frac{3}{4}\right)}\, _2F_1\left(\frac{1}{4},\frac{5}{4};\frac{5}{2};x\right)}{x^{1/4}}\, ,
\end{align}
and is analytic at $r=r_0$ ($x=1$) due to the cancellation of the branch cuts starting at $x=1$ that are separately present in the two hypergeometric functions.

It would be interesting to further analyze this solution. It seems like it is not possible to have a perturbation that decays towards the singularity $r\rightarrow 0$. Instead, the perturbation decays towards the horizon, i.e. it is decaying in tortoise coordinates. So the presence of the other throat appears to have a significant effect on the interior, where the perturbation becomes large again as we approach the singularity, since $Q(x\rightarrow \infty) \sim \frac{2 \sqrt{2 \pi } x}{\Gamma \left(-\frac{1}{4}\right) \Gamma \left(\frac{3}{4}\right)}$. The singularity inside this wormhole is therefore not of the AdS-Schwarzschild type. The $S^5$ does not factorize, so the geometry is really a full ten-dimensional wormhole.

\subsection{Global Structure and Flux Conservation}
\label{sec:global}

Here we discuss how regions I-II-III (the wormhole) should be glued to regions IV-V (the two throats in a single spacetime) so that the five form flux is conserved.\footnote{We thank Juan Maldacena for raising this point.} The gluing procedure leads to some interesting global properties of the wormhole. We will show that the spacetime has a moduli space coming from the freedom to add a certain amount of twisting during gluing.

\begin{figure}[hbtp!]
\begin{center}
\subfloat[\label{fig:wormholeflux}]{\includegraphics[width =.5\textwidth]{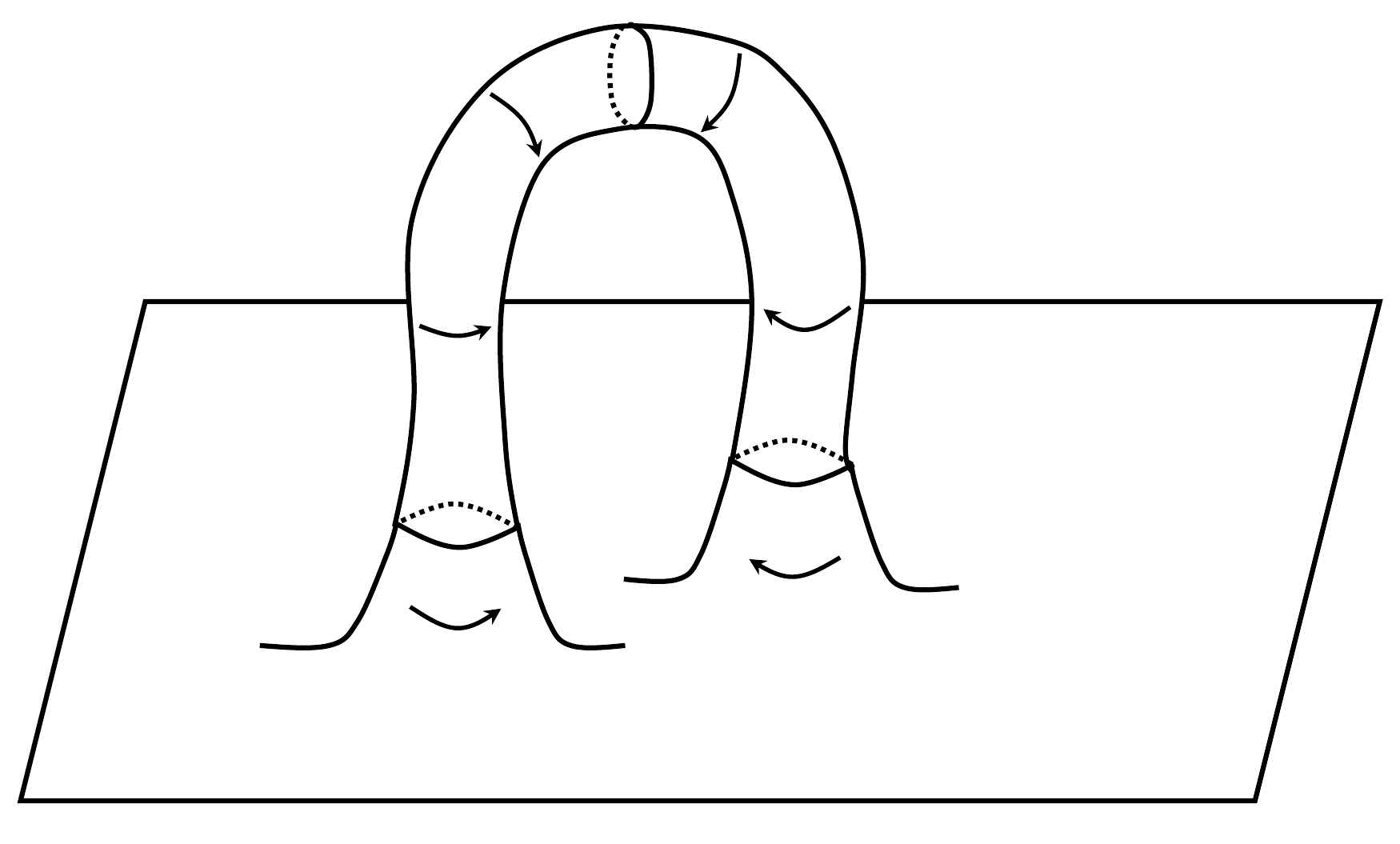}}
\subfloat[\label{fig:kleinflux}]{\includegraphics[width =.5\textwidth]{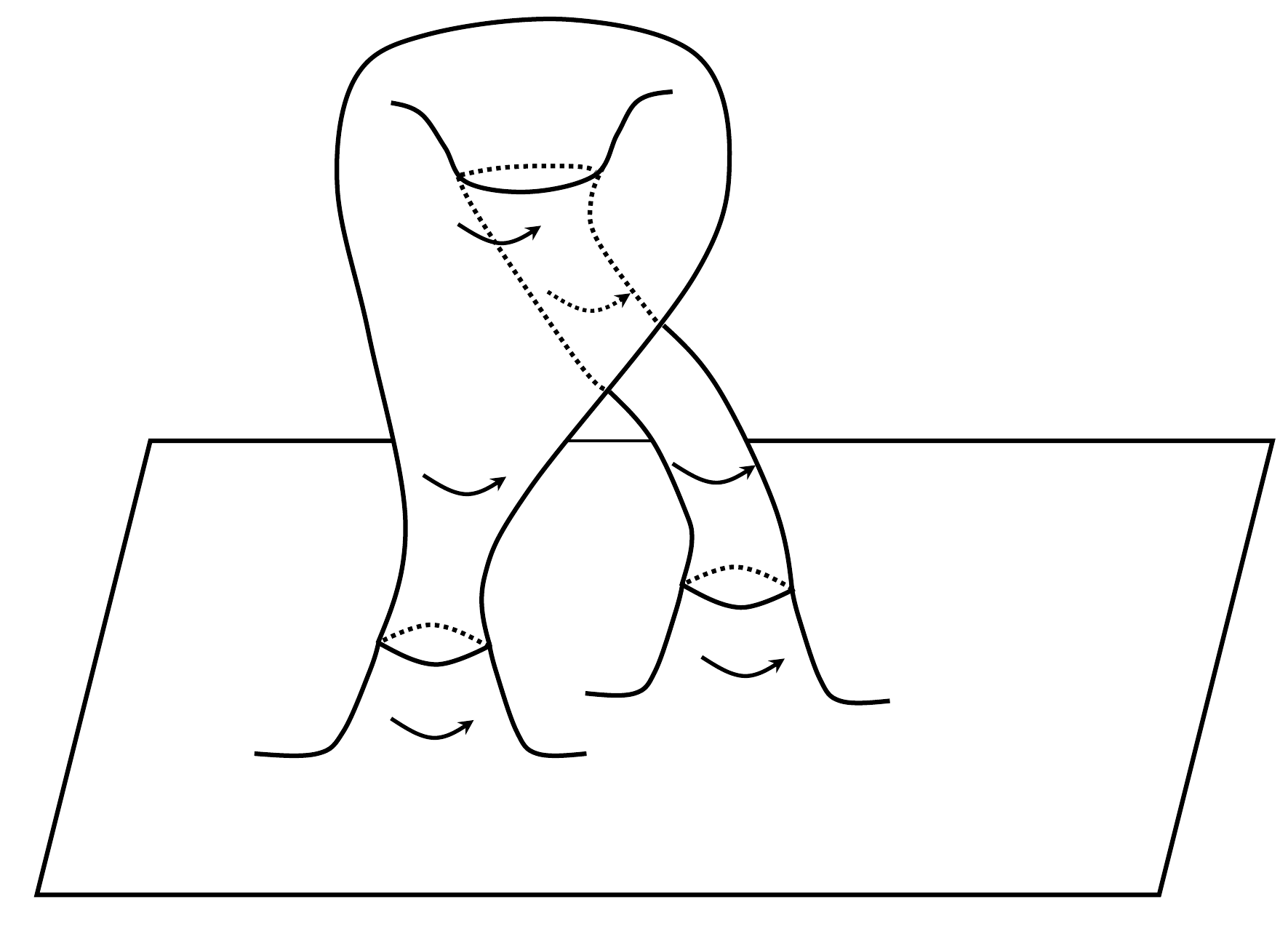}}
\end{center}
\caption{Spacetime wormholes with orientable (a) and non-orientable (b) Cauchy slices. In the orientable wormhole, conservation of circuital flux demands that the flux reverses direction at the second throat relative to the first throat, while in the non-orientable wormhole the flux at both throats points the same direction.\label{fig:fluxgeom}}
\end{figure}

In the geometry that we have described, both AdS throats have a positive net five-form flux towards infinity, so that there are $2N$ units of flux near infinity and $N$ near each throat. This presents a puzzle: if there are no sources in the wormhole, flux conservation demands that the flux should thread through the wormhole and close in the outside, giving zero net flux far away from the throats rather than $2N$. Our setup is analogous to the circuital law for a magnetic field in two dimensions, where the closed line integral must be conserved if there are no sources for the curl. This is illustrated for a 2D wormhole in Fig.~\ref{fig:wormholeflux}, where again, the wormhole without sources has zero line integral for the magnetic field on a loop enclosing both throats. However, one may support a non-vanishing line integral on such a loop purely by modification of the geometry, without adding sources. This is achieved by cutting open the wormhole, and gluing it back to the ambient space while twisting to invert the angular coordinate. This results in a sourceless ``Klein-bottle" wormhole, which is a non-orientable surface that supports a nonzero circuital flux at infinity. This is shown on Fig.~\ref{fig:kleinflux}.

\begin{figure}[hbtp!]
\begin{center}
\includegraphics[width=.7\textwidth]{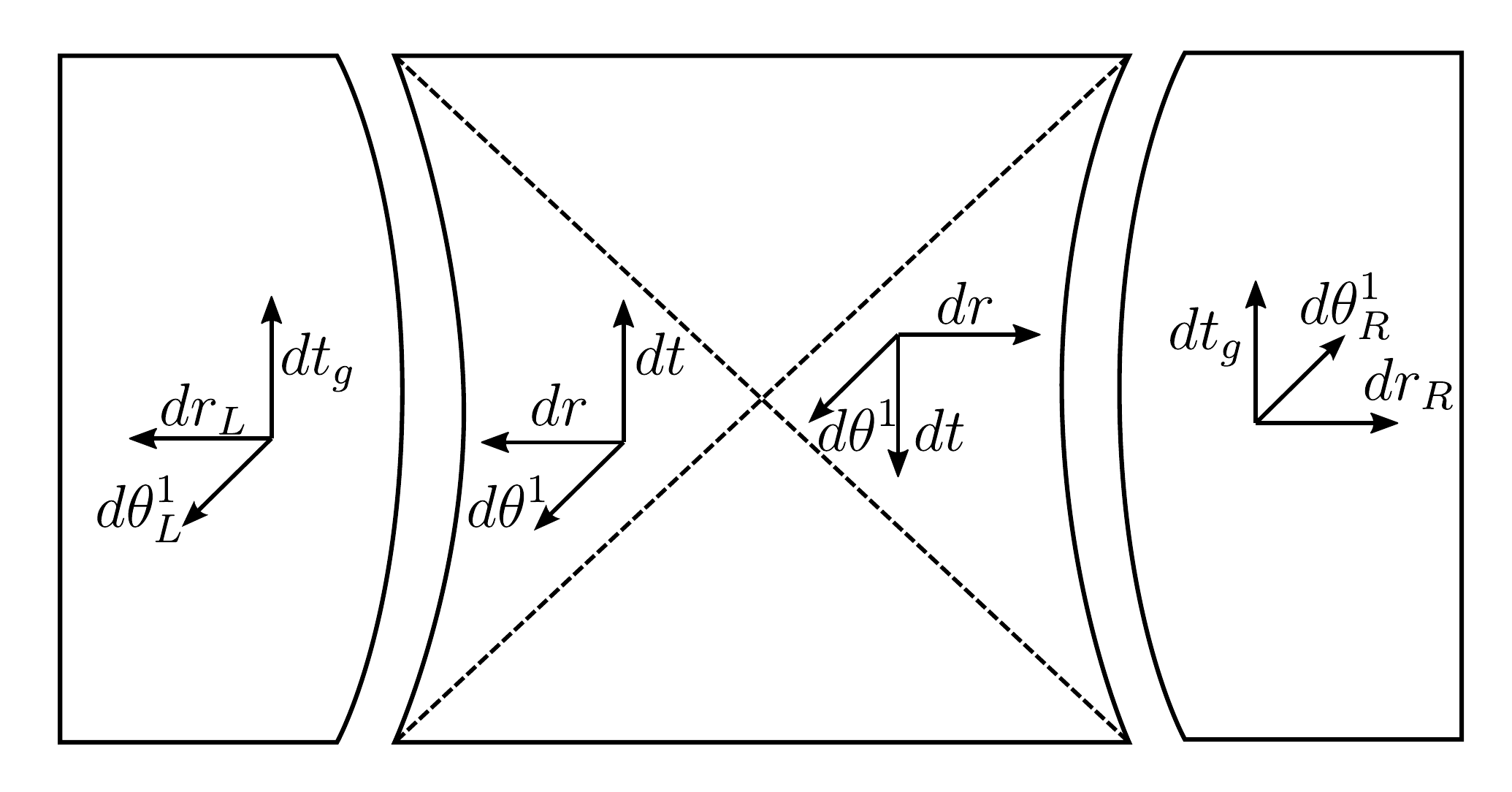}
\end{center}
\caption{Spacetime diagram of gluing the wormhole to the throat regions. In the middle we have the Penrose diagram of the eternal black brane, and the sides represent the throat regions. We show the orientation of the coordinate differentials $dr$, $dt$, and $d\theta^1$. The $d\theta^1$ differential points out from the plane of the figure in the left throat and in the wormhole, but it points inwards in the right throat. The three differentials must always form the same right handed system.\label{fig:penroseglueing} }
\end{figure}

The way that flux conservation works in our supergravity wormhole is very similar, although there are some technical differences because the flux comes from a five-form and lives in ten dimensions. In particular, the Cauchy slices of the wormhole will remain orientable. We illustrate on a spacetime diagram in Fig.~\ref{fig:penroseglueing} the two throats and the wormhole before we glue them together, and the orientation of the coordinate differentials. The left and right throats share a time coordinate $t_g$ and three spatial coordinates $x^i_g$ which are globally defined with the same orientation in the ambient space outside the throats. However, the natural radial and angular coordinates $r_L$ and $\theta_L^i$ at the left throat do not coincide with the corresponding coordinates $r_R$, $\theta_R^i$ at the right throat. This is because the geometry only fibers into AdS$_5 \times S^5$ near each throat, so the two five-spheres are centered at different points.

For the following discussion, by the \emph{electric part} of the five-form we refer to the term proportional to $dt\wedge dx^1\wedge dx^2 \wedge dx^3 \wedge dr$ and by the \emph{magnetic part} we refer to the term proportional to $d\text{Vol}_{S^5}$. Now, both electric and magnetic parts of the five-form are oriented in the same direction in the ambient spacetime, so deep in each throat, both have the same expression in local coordinates: $dt_g \wedge dx^1_g \wedge dx^2_g \wedge dx^3_g \wedge dr_{(L/R)}$ for the electric part and $d\text{Vol}_{S^5 (L/R)}$ for the magnetic part. At the left side of the wormhole, we choose the exterior Schwarzschild coordinates in the left wedge of the Penrose diagram to match the direction of local coordinates of the left throat: $dt = dt_g$, $dx^i = dx^i_g$, $dr = dr_L$, $d\theta_i = d\theta^i_L$ \footnote{The equalities that describe the ``gluing" between the left/right throats and left/right exterior wedges should be understood to be specifying the transition functions on the wormhole manifold in the coordinate patches where they are defined.}. In the right wedge, the radial coordinate points outwards towards the right throat, and the Schwarzschild time coordinate runs downward, in the opposite direction as the left wedge. However, we would like to glue the throats to the wormhole so that time points up on both sides. This is what we expect from the field theory, since after Higgsing the SYM Hamiltonian looks like $H_L + H_R$ in the IR, which generates upwards time evolution on both sides. Therefore, in the right wedge, we must take $-dt = dt_g$ and $dr  = dr_R$. 

The gluing of the rest of the coordinate directions at the right interface is determined by requiring the five form to be continuous. Consider starting with the five-form in the left throat and continuing into the left exterior wedge and across the wormhole to the right exterior wedge, where we must glue the geometry back to the right throat. On Fig.~\ref{fig:penroseglueing} we show the orientation of the coordinate differentials that are changing during this process. The rest of the coordinate differentials, $dx^i$, $i=1,2,3$ and $d\theta^i$, $i=2,\ldots ,5$ are oriented the same way throughout the figure. In the Schwarzschild coordinates, the five-form has the same solution in both the left and right wedges. But note that in the right exterior wedge of the wormhole, both $dr$ and $dt$ are flipped in Schwarzschild coordinates relative to the left exterior wedge.  Since both of these are flipped, the electric part of the five-form, $F_5 \sim dt\wedge dx^1\wedge dx^2 \wedge dx^3 \wedge dr$ actually keeps its orientation throughout the wormhole region. The same applies for the magnetic part, since the $S^5$ approximately factorizes in the wormhole. 

On the other hand, in the right throat, the basis differentials $dr_R$ and $d\theta^1_R$ are flipped relative to the left throat. However, the solution for the five-form looks the same in terms of these coordinate differentials in both throats.  So the orientation of \textit{both} the electric part $dt_R\wedge dx^1\wedge dx^2 \wedge dx^3 \wedge dr_R$ \textit{and} the magnetic part $d\theta^1_R \wedge d\theta^2\wedge d\theta^3 \wedge d\theta^4\wedge d\theta^5$ of the five form appear reversed compared to the right wedge of the wormhole for the purpose of gluing them.\footnote{What we mean here is that tracking the global five-form from the left throat to the right throat on the outside results in five-forms that point in opposite directions in the left and right throats if we draw them as in Fig.~\ref{fig:penroseglueing}.} So a direct gluing would lead to a discontinuous five-form. However, we can follow the idea from Fig. \ref{fig:kleinflux} and perform the gluing by twisting the $x^i$ coordinates (parallel to the brane) and the $\theta^1$ coordinate by an inversion at the gluing surface.

In terms of transition functions between the right wedge and the right throat, this works as follows. First we align the basis of coordinate differentials on the two sides by introducing new coordinates $t'=-t$ and $(\theta^1)'=-\theta^1$ in the right Schwarzschild wedge. In this properly aligned basis, there is an explicit sign difference in both the electric and magnetic components of the five form compared to the right throat. Then in order to make the five form components continuous, we glue with the transition functions $t_R\equiv t_g=t'$, $x_R^i\equiv x^i_g=-x^i$ and $\theta^1_R=-(\theta^1)'$. This way, in the electric part we make up for the sign by inverting $x^i$ in the gluing function, while in the magnetic part we invert the $\theta^1$ direction. This makes the complete five form continuous. Note that since the total determinant of this twist is positive, the Cauchy slice remains orientable, as opposed to the 2d example of Fig. \ref{fig:kleinflux}.\footnote{The total spacetime is also orientable since there exists a globally defined ``upwards" time, which in the wormhole region is just Kruskal time.}

Note that this twisted gluing results in a smooth geometry, because the $O(3)$ symmetry of the $x^i$ subspace is unbroken by the configuration of two throats, so nothing will depend on these coordinates even in the fully nonlinear time-dependent solution that we have not written down. Similarly, there is an unbroken $O(5)$ subgroup of the $O(6)$ acting on the $r_i$ coordinates, where the $O(5)$ fixes $r_1$, the direction in which the throats are separated. The $\theta^1$ twisting is an inversion of the $r_2,\ldots ,r_6$ coordinates, which is an element $R\in O(5)\subset O(6)$ of this unbroken symmetry with $\det R=-1$.

Now we briefly discuss the moduli space of solutions. Note that the only restrictions on the spatial twisting at the right gluing surface are that (i) it is from the subgroup which remains a symmetry of the solution, (ii) it reverses the orientation of both $dx^1\wedge dx^2 \wedge dx^3$ and $d\text{Vol}_{S^5}$. This gives a freedom in picking the group element with which we twist the gluing, resulting in a moduli space. In addition to the spatial twisting, as pointed out in \cite{Verlinde2020}, one may introduce a constant time shift in the identification of Schwarzschild time with the global time in the throat, which gives an extra real parameter (the difference between constant time shifts between left and right). Therefore, the total moduli space of single boundary wormholes is
\begin{equation}
    \mathbb{R} \times ISO(3) \times SO(5).
\end{equation}
Here, $ISO(3)$ denotes the group of (orientation-preserving) rotations and translations of the $x^i$ coordinates. 

Finally, let us comment on the field theory interpretation of the gluing twisted by inversions. The inversion of Schwarzschild time and the parallel coordinates $x^i$  correspond to time reversal and parity (TP) in the right IR $\mathcal{N}=4$ SYM factor. The inversion on the $S^5$ corresponds to inverting the $R$ charges in the field theory, so it is natural to think about it as the action of charge conjugation $C$. Therefore, from the field theory point of view, the right IR field theory factor is ``glued back" to the UV field theory by an action of CPT. This is  natural for the following reason. As discussed before, the state in the IR looks like the thermofield double state. The TFD state is defined from the square root of the thermal density matrix $\rho^{1/2}$, which is an element of $\mathcal{H}\otimes \mathcal{H}^*$, where $\mathcal{H}^*$ denotes the dual Hilbert space, where bra vectors live. In order to define the thermofield double, which lives on a doubled Hilbert space $\mathcal{H}\otimes \mathcal{H}$, one needs to turn the bra vectors into ket vectors with an anti-unitary symmetry. There is one such anti-unitary transformation that is a symmetry in any quantum field theory, which is CPT.

\section{Instability of the Solution}\label{sec:stability}

\subsection{Instability Timescale}

The wormhole solution we have described is not an extremal (BPS) solution of type IIB: its mass is larger than its charge. Consequently, it suffers from an instability: the attractive gravitational (NS-NS) force is larger than the repulsive five-form (R-R) force between the underlying branes, so that at late times the wormhole disappears as the two stacks of $N$ branes collide and form a single stack of $2N$ branes at nonzero temperature. We can compute the time scale of this instability by examining the tree-level effective action governing the dynamics of one stack of branes in the background of the other stack. The full action is the Dirac-Born-Infeld (DBI) action describing the geometric dynamics of the branes and their coupling to open strings, plus the coupling of the branes to the five-form \cite{GIBBONS1998603}:
\begin{align}
    S = -T \int d^{p+1} \xi \:\sqrt{-\text{det}\,G_{ab}} + \mu\int d^{p+1} \xi \: C_4 \, . \label{eq:dbiaction}
\end{align}
Here the $\xi^i$ are coordinates on the world-volume of a brane, $T$ is the tension of the stack of branes, $\mu$ is the charge density coupling to the five-form, $G_{ab}$ is the pullback of the background metric $g_{\mu \nu}$ to the brane, and $C_4$ is the pullback of the potential for the five-form. Evaluating the DBI action using the classical metric and five-form gives the effective action at tree level where we have taken the backgrounds for the antisymmetric two-form and the gauge field on the brane to be zero. 

In order to study the dynamics of one stack of branes as a probe, the backreaction of the probe on the background geometry should be negligible. However, this is not the case in the full two-center geometry, as each stack of branes sources its own independent AdS throat. Moreover, each of the stacks have a field that is the size of the AdS radius, and since they are separated in the asymptotic $S^5$ directions which have comparable size, we cannot treat the two stacks as point-like objects interacting via weak fields. Nevertheless, we may obtain a lower bound on the timescale of the instability by considering the motion of an extremal probe brane located halfway between the two stacks of branes, where we heat up one stack slightly and leave the other extremal. We may think about the extremal probe as being separated from the extremal stack. Such a brane experiences a higher acceleration than a brane located deep in the AdS throat of the extremal stack, where it is further from the thermal stack. We will show that the temperature and separation of the branes can be chosen so as to make this lower bound on the instability timescale arbitrarily high, i.e. the wormhole is long-lived.

The extremal probe brane starts at a point in the geometry which can be approximated by the flat-space region of \eqref{eq:metricnonpert}, \eqref{eq:fiveformnonpert}, far from both the horizon of the thermal branes and the AdS throat of the extremal branes. To compute the pullback of the metric and four-potential, we use spacetime Lorentz transformations and world-volume reparameterizations to work in ``static gauge" in which the world-volume coordinates are parallel to the spacetime coordinates.
\begin{align}
\xi^0 = t, \qquad \xi^i = x^i \, ,
\end{align}
where $i = 1,2,3$. The pullback of the potential to a brane sitting at distance $r$ from the thermal stack is then
\begin{align}
    C_4 = \sqrt{1+\frac{r_0^4}{\Lambda^4}} \left(\frac{L^4}{r^4} + \frac{L^4}{\Lambda^4} \right)^{-1} dt \wedge dx^1 \wedge dx^2 \wedge dx^3 \, .
\end{align}
In spherical coordinates, the probe brane moves only in the radial direction, so the pullback of the metric is given by 
\begin{align}
    G_{00} = g_{00} + \dot{r}^2 g_{rr}, \qquad G_{ii} = g_{ii} \, ,
\end{align}
and all other components are zero. Consequently, the effective action experienced by the probe brane is, defining $M = TV$ and $Q = \mu V$ as the effective mass and charge of the probe brane where $V = \int d^3\xi^i$ is the (regularized) brane world-volume,
\begin{align}
    S = \frac{\Lambda^2}{L^4}\int dt\, \frac{r^4}{r^4+\Lambda^4}\biggl( -M \Lambda^2 \sqrt{1-\frac{r_0^4}{r^4}- \dot{r}^2 \frac{L^4(r^4+\Lambda^4)}{\Lambda^4(r^4-r_0^4)}} + Q\sqrt{r_0^4+\Lambda^4}\biggr) \, .
\end{align}
This expression should be expanded at large radius compared to the horizon $r_0$, but keeping $r/\Lambda$ fixed since $r \sim \mathcal{O}(\Lambda)$ at the scale of the dynamics. Therefore, we introduce the dimensionless radial coordinate $\hat{r} = r/\Lambda$ that we imagine to be order one, and dimensionless time $\hat{t} = \frac{r_0 t}{L^2}$ as before. Since we take the probe to be extremal, we set $Q=M$, and all dimensionful quantities then scale out in front of the action,
\begin{align}
    S = \frac{M \Lambda^4}{r_0 L^2} \int d\hat{t} \,\frac{\hat{r}^4}{\hat{r}^4+1}\biggl(\sqrt{1+\epsilon^4} - \sqrt{1-\frac{\epsilon^4}{\hat{r}^4}-\epsilon^2 \left(\frac{d\hat{r}}{d\hat{t}} \right)^2 \frac{\hat{r}^4+1}{\hat{r}^4-\epsilon^4}} \biggr) \, ,
\end{align}
where $\epsilon=r_0/\Lambda$ as before. We may now expand in $\epsilon$ and take $d\hat{r} / d\hat{t}\ll 1$, the Newtonian slow-moving approximation for the probe\footnote{The speed $d\hat{r} / d\hat{t}$ is $\mathcal{O}(\epsilon^3)$ in this expansion, coming from balancing the orders of the leading potential and kinetic terms and/or from the equation of motion.}, to find at lowest order
\begin{align}
    S = \frac{M\Lambda^2 r_0}{ L^2} \int d\hat{t} \left(\frac12  \left(\frac{d\hat{r}}{d\hat{t}}\right)^2 + \frac{\epsilon^2}{2} + \frac{\epsilon^6}{8} \left(\frac{1}{\hat{r}^4} - 1\right) + \ldots \right) \, .
\end{align}
This is motion in a flat space attractive Coulomb potential which scales as $\mathcal{O}(\hat{r}^{-4})$ as expected for a charged object of codimension six in ten spacetime dimensions, in agreement with what would have been found from the tree-level closed string exchange. The resulting dynamics are simply
\begin{align}
    \frac{d^2 \hat{r}}{d\hat{t}^2} = -\frac12 \frac{\epsilon^6}{\hat{r}^5} \, ,
\end{align}
and so the acceleration can be made small by making $\epsilon=r_0/\Lambda$ small. In terms of the original time coordinate, the instability timescale is $t \sim \frac{L^2 \Lambda^3}{r_0^4}\sim \beta \epsilon^{-3}$, where $\beta=\frac{\pi L^2}{r_0}$ is the inverse temperature of the black brane. That is, taking the thermal branes to be very cold or the stacks of branes to be widely separated, the wormhole solution can be made arbitrarily long-lived.

\subsection{Stabilizing with Rotation}
\label{sec:rotation}

One may wonder whether our wormhole can be stabilized by making the throats spin around each other in the transverse $r_i$ directions. We will not attempt to perturbatively construct such a spinning solution in the present work. On the other hand, we can repeat the DBI analysis above for the case where the extremal probe brane rotates around the non-extremal black branes. We parameterize the brane trajectory in a circular orbit around the equator $\theta_1=\ldots=\theta_4=\pi/2$ by $r(t)$, $\theta_5(t)$. In this case, the $G_{00}$ component of the pullback of the metric is
\begin{align}
    G_{00} = g_{00} + \dot{r}^2 g_{rr} +  \dot{\theta}_5^2 g_{\theta_5 \theta_5} \, .
\end{align}
Using the same coordinates as the previous section, the DBI action for the extremal probe constrained to the equator is
\begin{align}
    S = \frac{M\Lambda^4}{r_0 L^4} \int d\hat{t} \frac{\hat{r}^4}{\hat{r}^4+1} \left(\sqrt{1+\epsilon^4} - \sqrt{1-\frac{\epsilon^4}{\hat{r}^4}-\epsilon^2 (\hat{r}^4+1) \left[\frac{1}{\hat{r}^4-\epsilon^4} \left(\frac{d\hat{r}}{d\hat{t}} \right)^2+\frac{1}{\hat{r}^2} \left(\frac{d\theta_5}{d\hat{t}} \right)^2 \right]} \right) \, . \label{eq:dbirotate}
\end{align}
Expanding in $\epsilon$ and taking the slow-moving approximation yields
\begin{align}
    S = \frac{M\Lambda^2 r_0}{ L^2} \int d\hat{t} \left(\frac12  \left(\frac{d\hat{r}}{d\hat{t}}\right)^2 + \frac12 {\hat r}^2 \left(\frac{d\theta_5}{d\hat{t}} \right)^2+ \frac{\epsilon^2}{2} + \frac{\epsilon^6}{8} \left(\frac{1}{\hat{r}^4} - 1\right) + \ldots \right) \, ,
\end{align}
the same result as previously with the Newtonian rotational kinetic energy added. The radial equation of motion, assuming the existence of a solution with a constant rotational velocity $\frac{d\theta_5}{d\hat{t}} = \hat{\omega}$, is 
\begin{align}
    \frac{d^2 \hat{r}}{d\hat{t}^2} = \hat{r} \omega^2 - \frac{\epsilon^6}{2\hat{r}^5} \, ,
\end{align}
and therefore we can obtain circular orbits of radius $\hat{r}_c$ when the rotation speed is
\begin{align}
    \hat{\omega} = \frac{\epsilon^3}{\sqrt{2} \hat{r}_c^3} \, .
\end{align}
Therefore, the angular speed needed to obtain circular orbits is $\omega \sim \epsilon^3/\beta$, the inverse of the instability time scale. In dimensionful coordinates, this speed is
\begin{align}
    \omega = \frac{r_0^4}{\sqrt{2} L^2 r_c^3} \, .
\end{align}
We can check if the circular orbit radius $\hat{r}_c = \frac{\epsilon}{(\sqrt{2} \hat{\omega})^{1/3}}$ leads to stable or unstable orbits. For this, we examine the effective potential written in terms of conserved angular momentum $\ell =  \frac{{\hat r}^2}{\epsilon^3} \frac{d \theta_5}{d\hat t}$,
\begin{equation}
    V(r)=\epsilon^6 \left( \frac{1}{4}-\frac{1}{4 \hat r^4}+\frac{\ell^2}{\hat r^2} \right) \, .
\end{equation}
We see that the circular orbit corresponds to a maximum, i.e. it is unstable. The reason this happens is that the centrifugal piece in the effective potential dies off slower than the attractive force, which is the opposite of the situation in normal 4D Kepler motion. Based on this analysis, it is unlikely that the wormhole solution can be stabilized by rotation, unless nonlinear effects conspire to stabilize a circular orbit.
    
We can also try to solve \eqref{eq:dbirotate} for circular orbit frequencies directly without series expanding by taking the circular orbit as an ansatz. In that case the equation of motion reduces to the algebraic equation
\begin{align}
    \hat{r}^{10} \omega^2 \epsilon^8+4 \hat{r}^6 \omega^2 \epsilon^8+3 \hat{r}^2 \omega^2 \epsilon^8+\hat{r}^4 \left(4 \sqrt{-\frac{\left(\epsilon^4+1\right) \left(\hat{r}^6 \omega^2 \epsilon^8-\hat{r}^4+r^2 \omega^2 \epsilon^8+\epsilon^4\right)}{\hat{r}^4}}-2 \epsilon^4-4\right)+2 \epsilon^4 = 0 \, .
\end{align}

We find an additional solution in this case with angular speed at leading order in $\epsilon$ given by
\begin{align}
    \hat{\omega} = \frac{2\sqrt{2}\hat{r}_c}{(\hat{r}_c^4+3) \epsilon} \, ,
\end{align}
or $\omega = \frac{\Lambda^4}{L^2} \frac{2\sqrt{2} r_c }{(r_c^4 + \Lambda^4)}$ in dimensionful coordinates. This solution did not appear previously from perturbing the action around small $\epsilon$ simply because it is inversely proportional to $\epsilon$ and therefore not perturbatively slow-moving\footnote{One might worry that $\omega$ exceeds light-speed, even if it does not diverge. One can check that the maximum value of $\omega$ is $\frac{3^{3/4} \Lambda}{\sqrt{2}L^2}$ which is certainly small as $L \gg \Lambda$, and that this occurs at the reasonable radius $r_c = \Lambda / 3^{1/4}$.}. This value of $\hat{\omega}$ supports two different possible radii,
\begin{align}
    \hat{r}_c = \frac{3\hat{\omega} \epsilon}{2\sqrt{2}} \qquad \text{and} \qquad \hat{r}_c = \frac{\sqrt{2}}{(\hat{\omega} \epsilon)^{1/3}} \, ,
\end{align}
or in dimensionful coordinates,
\begin{align}
    r_c = \frac{3L^2 \omega}{2\sqrt{2}} \qquad \text{and} \qquad r_c= \sqrt{2} \left(\frac{\Lambda^4}{\omega L^2}  \right)^{1/3} \, .
\end{align}
We can study the stability of these circular orbits by linearizing around the solution. One finds that the (dimensionless) frequency-squared of the radial oscillation, to leading order in $\epsilon$ is
\begin{equation}
    \Omega^2 = \frac{4 \left(5 \hat{r}_c^8+12 \hat{r}_c^4-9\right) \hat{r}_c^2 \left| \hat{r}_c^4-1\right| +4 \left(\hat{r}_c^{12}-5 \hat{r}_c^8-33 \hat{r}_c^4+5\right) \hat{r}_c^2}{\epsilon^2 \left({\hat r}_c^4+1\right)^3 \left(\hat{r}_c^4+3\right)^2} \, ,
\end{equation}
that is, the orbit is stable when ${\hat r}_c\gtrsim 1.281$ and unstable when ${\hat r}_c\lesssim 1.281$. In the extremal limit, these circular orbits rotate with a finite angular velocity, as $\omega$ does not depend on $r_0$.  These orbits are not directly relevant for stabilizing our wormhole, which is a perturbation to a non-rotating solution. This is because for self-consistency we would want the rotation in the circular orbits to be perturbatively small in $\epsilon=r_0/\Lambda$, while we saw that the rotation persists even in the extremal limit.   One may therefore wonder if there exists a rotating version of the extremal two-center solution that is perturbatively stable. An exact solution is likely not possible  due to gravitational and five form radiation, but we really just want a long-lived rotating binary black hole. This would provide a starting point for a wormhole solution stabilized by rotation. 

\section{Traversing the Wormhole} \label{sec:traverse}

The two throats in our wormhole are separated by causal horizons, so it is not possible to traverse through it. Near the horizons, the wormhole looks like a perturbation of the planar AdS-Schwarzschild black brane, which is a marginally non-traversable solution in the sense that it can be made traversable by a small negative energy perturbation \cite{GaoJafferisWall}. Here we wish to analyse if the perturbation of the geometry near the horizon spoils this property. In two-sided null Kruskal coordinates $U,V$ (which exist for both the eternal black brane and for our wormhole geometry) the requirement to violate the ANEC is written $\int dU \, T_{UU} < 0$ along $V = 0$. In the absence of any stress-energy, $T_{UU} = 0$, null rays along $V = 0$ pass through the bifurcation surface and asymptote to infinity in either direction. Consequently, any negative perturbation will pull back the horizons and create traversability.

We will now evaluate the ANEC for the monopole- and dipole-corrected Einstein-Rosen bridges of Sec.~\ref{sec:monopole} and Sec.~\ref{sec:dipole}. The monopole corrections \eqref{eq:monopole} do not affect the marginal traversability since  $T_{UU}$ vanishes along the horizon. This follows because as noted at the end of Sec.~\ref{sec:monopole}, this correction can be obtained by linearizing \eqref{eq:metricnonpert}, which is the asymptotically flat black brane in rescaled coordinates. The dipole contribution \eqref{eq:dipole} is more complicated to analyze because the $t-r$ plane is no longer decoupled from the $\theta_1$ angle, so the near-horizon geometry is effectively three-dimensional. Regardless, the location of the horizon stays at $r=r_0$ since the location of the zero of the blackening factor is not affected. Moreover, the null geodesics comprising the horizon remain on the $t-r$ plane at fixed $\theta_1$. This can be seen by examining the $\theta_1$ component of the geodesic equation,
\begin{equation}
 \frac{d}{d\lambda}   (r^2\dot{\theta_1}) = -\frac{r^2 \sin \left(\theta _1\right) h_g(r) \left(\dot{r}^2 L^4 r^4+\dot{t}^2 \left(r^4-r_0^4\right){}^2\right)}{L^2 \Lambda^5 \left(r^4-r_0^4\right)} \, ,
\end{equation}
where dot indicates derivative with respect to affine parameter $\lambda$. We have $h_g(r)=\frac{20 \pi ^{3/2} (r-r_0)}{\Gamma \left(-\frac{1}{4}\right)^2} + \cdots$ around $r=r_0$, so in order to have $\ddot{\theta_1}=0$ at the horizon, we need that $\dot{r}$ vanishes at $r=r_0$.  Examining the condition $g_{mn}\dot{x^a}\dot{x^b}=0$ around $r=r_0$ one finds that $\dot{r} \propto \sqrt{r-r_0}\dot{\theta_1}$. Therefore, $\dot{r}=\dot{\theta_1}=0$ and $\dot{t}= \text{const}$ is a null geodesic at $r=r_0$ for any fixed $\theta_1$ (and the rest of the seven coordinates fixed as well). Therefore, $t$ also affinely parameterizes the null worldlines, so the ANEC quantity can be written $\int dt\, T_{tt}$. We identify the stress-energy tensor from the right-hand side of \eqref{eq:geomEOM2} as
\begin{align}
    T_{\mu \nu} = \frac{1}{4\cdot 4!} F_{\mu \alpha \beta \gamma \delta} F_{\nu}^{\:\:\alpha \beta \gamma \delta} \, .
\end{align}
Applying this to the solution \eqref{eq:dipole} we find that $T_{tt}=0$ at $r=r_0$ up to $O(1/\Lambda^6)$ corrections. Therefore, at the order $\mathcal{O}(\Lambda^{-5})$ of the dipole corrections, spherical symmetry breaking does not affect the marginal non-traversability of the single-boundary wormhole.

As in \cite{GaoJafferisWall, SYKwormhole}, one mechanism to generate negative contributions to the ANEC that allow traversability is to introduce a nonlocal coupling between the two throats of the wormhole by adding a double-trace type interaction in the field theory. In fact, the field theory symmetry breaking $SU(2N) \to S(U(N) \times U(N))$ that we have described already generates couplings between the two $SU(N)$ effective subfactors in the IR from the Wilsonian RG flow \cite{INTRILIGATOR200099, connectivity}. At leading order, these include single-trace interactions of the form $g_I V_I$, where $V_I$ is proportional to
\begin{align}
V_I \propto \text{tr} \left(F_{\mu \nu} F^{\nu \rho} F_{\rho \sigma} F^{\sigma \mu} - \frac14 (F_{\mu \nu} F^{\mu \nu})^2 \right) \, ,
\end{align}
and $I=1,2$ are the two $U(N)$ factors. The couplings $g_I$ are dynamically determined by abelian singleton degrees of freedom in the other CFT factors (i.e., the Goldstone modes associated to the moduli of branes in the other stack(s)). Of more interest to us with respect to traversability are the double-trace interactions that are generated.  These directly couple the IR factors in the CFT:
\begin{align}
    V_{IJ} \propto \text{tr}_I \left(F_{\mu \nu} F^{\mu \nu} \right) \text{tr}_J \left(F_{\mu \nu} F^{\mu \nu}  \right) \, .\label{eq:symdoubletrace}
\end{align}
When the full UV CFT is genuinely a product of $n$ individual subfactors $\text{CFT} = \prod_{i=1}^n \text{CFT}_i$, the dual bulk geometry generally consists of $n$ different asymptotic universes. In this case, the single-trace terms $\text{tr}_I F^2$ that comprise the operator \eqref{eq:symdoubletrace} are dual to the bulk dilaton in component $I$ \cite{OGwitten, PhysRevD.59.104021}. In our setting, the components $\text{tr}_1 F^2$ and $\text{tr}_2 F^2$ are dual to the bulk dilaton in the vicinity of the first and second throats, as these deep bulk regions correspond to the IR of the CFT where the approximate factorization into two $SU(N)$ gauge theories holds. Therefore, the double-trace interactions $V_{12}$ are structurally of the form $h_{12} \phi_1 \phi_2$ required to generate negative contributions to the ANEC as shown in \cite{GaoJafferisWall}. This indicates that the natural operators that arise from the Wilsonian RG flow in the IR of the symmetry-broken theory are of the correct form to generate traversability, albeit possibly weak traversability. However, in the Gao-Jafferis-Wall protocol, only one sign results in a traversable wormhole, while the opposite sign lengthens the wormhole.  It would therefore be interesting to determine the sign of the coefficient of \eqref{eq:symdoubletrace} as generated by the Wilsonian RG, at least in perturbation theory.  In fact, in our setting, there are various other double-trace operators that can be generated by the supersymmetry transformations of \eqref{eq:symdoubletrace}. A full analysis should understand the net effect of all such RG-generated double-trace operators on the sign of the null stress-energy. 

In \cite{GaoJafferisWall}, the double-trace interactions are taken to be relevant deformations of the Hamiltonian so that they are renormalizable and there is no backreaction at the AdS boundary. The term $V_{12}$ generated by the RG flow is an irrelevant deformation; nonetheless, this is not a concern as we know that the theory is UV-complete, since above the Higgs scale it flows to the $SU(2N)$ $\mathcal{N}=4$ SYM theory. Furthermore, \cite{GaoJafferisWall} take the deformation to be a quench, turned on after some time $t_0$. Since our solution is perturbatively unstable, we also expect the coupling strength to be time-dependent, although we have not analyzed this in detail. Lastly, \cite{GaoJafferisWall} takes the boundaries to be connected with the same time orientation by taking the deformation to be structurally $h(t) \phi_1 (t, \vec{x}) \phi_2 (-t, \vec{x})$. This is because the asymptotic time on one boundary of the eternal black hole runs in the opposite direction on the other boundary. In our setting, the wormhole resides in a single universe and we have taken time to run upwards on both sides, so there is a unique asymptotic time $t$ and we need not flip the time orientation between the two throats.

In addition to the terms that are naturally generated by RG-flow, we can try, like \cite{GaoJafferisWall}, to add by hand some deformation that generates traversability in the IR wormhole. This should be a relevant operator in order for it not to destroy the UV $SU(2N)$ $\mathcal{N}=4$ SYM theory. The lightest single trace operators in a single factor of $SU(N)$ $\mathcal{N}=4$ SYM are the $\Delta=2$ scalars in the $\textbf{20}$ of the $SO(6)$ $R$-symmetry. They are of the form $\mathcal{O}^{ij}=\text{Tr} \phi^{(i}\phi^{j)}$. The possible deformations $\mathcal{O}^{ij}_L \mathcal{O}^{kl}_R$ therefore furnish $\textbf{20}\times \textbf{20}$. These are marginal to leading order in $1/N$ due to large $N$ factorization. In order to work out the effects of deforming by these operators (with either sign of the coefficient) we would need to understand their RG flow and the $1/N$ corrections to their dimension. The corresponding single-sided double-trace operators (i.e. an operator in one of the low energy $SU(N)$ factors) $\mathcal{O}^{ij}_L \mathcal{O}^{kl}_L$ are well understood in the strong coupling regime \cite{Arutyunov:2000ku}, and they all have either vanishing or negative anomalous dimensions. The negative anomalous dimensions are intuitively understood as binding energies coming from the attractive nature of the bulk interaction between two particles. The same intuitive reasoning applies to the two-sided operator (i.e. an operator connecting the two low energy $SU(N)$ factors) $\mathcal{O}^{ij}_L \mathcal{O}^{kl}_R$, which suggests that these operators should be marginally relevant at strong coupling, and one should be able to use them to make our single boundary wormhole traversable.

\section{A Double Wormhole Between Universes} \label{sec:doublewormhole}

\begin{figure}[hbtp!]
\begin{center}
\includegraphics[width=.45\textwidth]{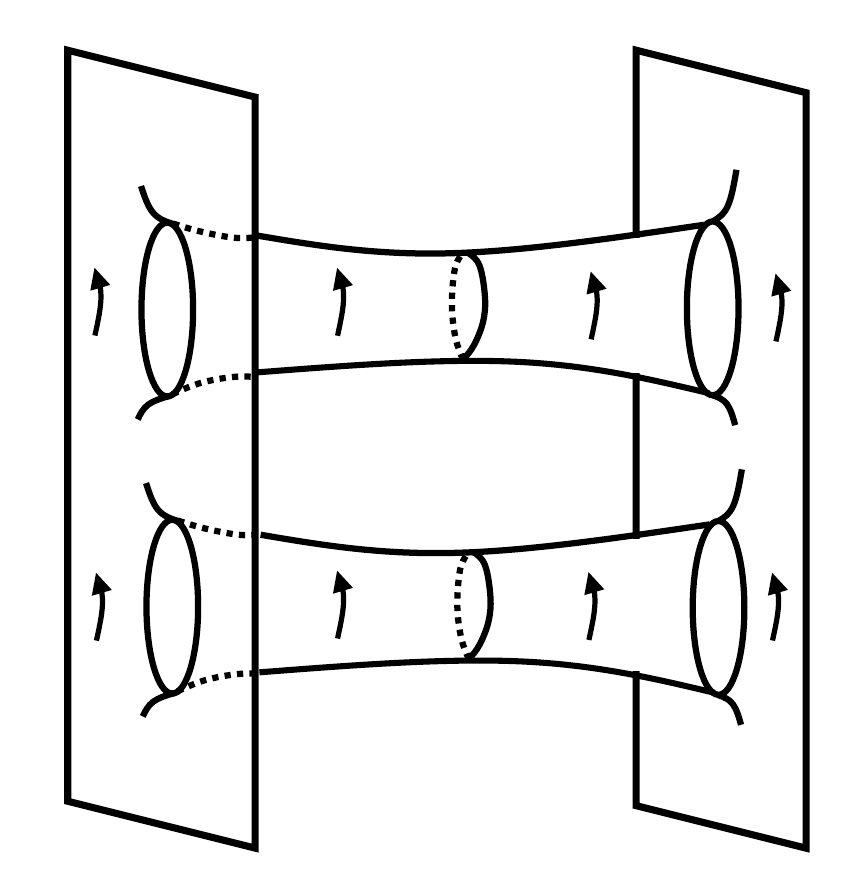}
\end{center}
\caption{A double wormhole between two asymptotically AdS universes.  This geometry, which can be constructed from the solutions in the text, is dual to a pair of Higgsed Yang-Mills theories, with IR factors entangled pairwise between them.
\label{fig:twouniverse}}
\end{figure}

The solutions we discussed in Sec.~\ref{sec:wormholeGeomSoln} capture certain effects of two non-extremal throats living in a single asymptotically AdS spacetime, and in  Sec.~\ref{sec:global} we explained how to join these throats so that we end up with a wormhole in a single universe. There are also other ways to join the solutions of Sec.~\ref{sec:wormholeGeomSoln} to get interesting new wormhole configurations. For example, one could duplicate the spacetime with two throats and join them in a way shown in Fig.~\ref{fig:twouniverse}. In this case, the global time runs in opposite way in the two asymptotic regions and no twisting is required to enforce flux conservation (see Fig.~\ref{fig:twouniverse}). This spacetime is patch-wise described by the same solutions that we have discussed in Sec.~\ref{sec:wormholeGeomSoln}, but the patches are glued together differently.

In the dual field theory we now start with two copies of $\mathcal{N}=4$ $SU(2N)$ SYM, and we Higgs each copy. Let us label the two theories $A$ and $B$, while the low energy factors are called $L$ and $R$. Then, the low energy Hilbert space is
\begin{equation}
    \mathcal{H}_{A,L}\otimes \mathcal{H}_{A,R} \otimes  \mathcal{H}_{B,L}\otimes \mathcal{H}_{B,R} ,
\end{equation}
and we expect a wormhole configuration like Fig. \ref{fig:twouniverse} to be approximately dual in the IR to a tensor product of two thermofield double states
\begin{equation}
    |\text{TFD}\rangle_{A,L;B,L}\otimes |\text{TFD}\rangle_{A,R;B,R}.
\end{equation}
We can embed this state in the UV Hilbert space $\mathcal{H}_A\otimes \mathcal{H}_B$ as explained in Sec.~\ref{sec:fieldtheory}, that is, we must take the temperatures of the thermofield doubles to be much smaller than the Higgs scale.

\section{Discussion}

In this paper, we  constructed an asymptotically AdS$_5 \times S^5$ single boundary wormhole solution by matching a two-center extremal black brane solution to a two-sided AdS black brane in perturbation theory.  Preserving continuity of the five-form in the solution required a global monodromy in some of the coordinates, although the total geometry remains orientable. The small parameter in the problem is the horizon radius compared to the separation of the throats, $r_0/\Lambda$. We argued that the solution is dual to an approximate thermofield double state in a single copy of $\mathcal{N}=4$ SYM, where the gauge group is Higgsed into two copies of $SU(N)$, which are entangled. In the field theory the small parameter is the ratio of the thermal scale to the Higgs scale. 

\subsubsection*{Thermal effective potential and $R$ charge}

Our wormhole is non-extremal, and consequently has to be unstable. This instability is dual in $\mathcal{N}=4$ SYM to the scalar vevs developing an effective potential at finite temperature, as illustrated in Fig. \ref{fig:effpotential}. We have argued that the wormhole can be made parametrically long lived by making $r_0/\Lambda$ small. 

Another possibility is to stabilize the wormhole by making the throats rotate around each other. In Sec.~\ref{sec:rotation} we  found that an extremal probe brane can be put on a stable circular orbit around a non-extremal black brane. This is surprising since planetary orbits are unstable in more than four dimensions, and is possible here due to the five-form interaction. The stable orbit we find has finite angular velocity in the extremal limit, so it is not possible to add this effect perturbatively to our solution. Nevertheless, this finding suggests that in the dual theory one can create a local minimum in the effective potential of the scalar vevs away from the origin by adding $R$ charge. This should lead to long-lived states with finite temperature symmetry breaking. The states are only long-lived, since from the supergravity picture, we expect them to decay due to gravitational and five form radiation. This is consistent with the expectation that all symmetries must be restored at sufficiently high temperatures: see \cite{Chai:2020zgq} for a recent discussion in the case of global symmetries. It would  be interesting to understand this effect better.

\subsubsection*{Global monodromy and moduli space}

As emphasized in \cite{Verlinde2020}, gluing the two sides of a wormhole to a single asymptotic region breaks the two-sided boost-like Killing symmetry of the eternal black brane geometry and correspondingly, there is a one-parameter family of wormholes labeled by the ``monodromy" of Schwarzschild time as one goes between the two throats on the outside. In addition to this, we have found that there is a freedom of introducing a global monodromy consisting of rotating and translating the parallel spatial directions to the brane, and also rotating by the unbroken $SO(5)$ subgroup of the $SO(6)$ symmetry of the $S^5$. 
Therefore, there is a moduli space $\mathbb{R}\times ISO(3)\times SO(5)$ of locally equivalent but globally different solutions. It would be interesting to understand the interpretation of this in the dual $\mathcal{N}=4$ $SU(2N)$ SYM theory. It is tempting to speculate that it is related to some ambiguity in embedding the IR state \eqref{eq:tfdapprox} into the UV theory, which possibly includes an ambiguity in the implementation of the energy cutoff in the state \eqref{eq:tfdapprox}.

\subsubsection*{Making the wormhole traversable}

We have showed that corrections coming from the two throats being in the same spacetime in the first few orders in perturbation theory do not spoil the marginal traversability of the wormhole, in the sense that the ANEC quantity $\int dU T_{UU}$ remains zero along the causal horizons. It would thus be interesting to see if the wormhole can be made traversable using the ideas in \cite{GaoJafferisWall}. This requires a double trace coupling between the two $SU(N)$ factors in the Higgsed $\mathcal{N}=4$ SYM theory. 

We have pointed out that such couplings are naturally generated in RG due to the fact that in the UV the two $SU(N)$ factors are part of the total $SU(2N)$. It would require a careful analysis to account for the net effect of all these double trace interactions and to see if the resulting sign makes the wormhole traversable. This is beyond the scope of the present paper but is certainly an interesting problem.

One may also try to make the wormhole traversable by adding a double trace coupling by hand. This would have to be a relevant double trace operator, otherwise the theory will no longer flow to a single $SU(2N)$ $\mathcal{N}=4$ SYM in the UV (or to a wormhole in a single spacetime). We have argued that such relevant double traces can be formed from the $\Delta=2$ scalar operators of the theory, though it would also be useful to check that the two-sided operators $\mathcal{O}_L^{ij} \mathcal{O}_R^{kl}$ have negative anomalous dimensions.

In \cite{SM_wormhole} negative contributions to the ANEC were generated by negative Casimir-like vacuum energies coming from the lowest Landau levels of the bulk fermion running in a cycle threading their wormhole solution. In our solution, there are various fermions in the spectrum of type IIB supergravity which have vacuum fluctuations, though we have set their classical backgrounds to vanish.  These fermions, and the bulk bosonic fields, should similarly provide Casimir-like vacuum energies in our setup. The sign of the total Casimir energy is important, as before; so it is important to check which contributions ultimately win out. There is potentially the possibility that the underlying supersymmetry enforces a vanishing total Casimir energy. In any case, the vacuum energies provide another potential mechanism for traversability in competition or collusion with the other effects that we have discussed.

\subsubsection*{Probing the monodromy through the wormhole}  

As we discussed, continuity of the five-form requires a twisted gluing of the interior AdS-Schwarzschild geometry to the two-center ambient spacetime, although the complete spacetime remains orientable. An interesting way of probing the resulting monodromy is to send a giant graviton through the wormhole. Giant gravitons are spherical D3-branes localized on the $S^5$ in the geometry, and are supported by their angular momentum and by interactions with the five-form flux \cite{Myers:1999ps,McGreevy:2000cw}. These brane states are created by determinant and subdeterminant operators in the field theory \cite{Balasubramanian:2001nh,Corley:2001zk,Balasubramanian:2004nb}. To use these branes to probe the wormhole in the field theory, we would want to construct such operators in the light infrared factors after Higgsing.  On the gravitational side, we could explicitly test what happens to the corresponding giant gravitons as they are moved through the wormhole, expecting them to emerge with inverted $\theta^1$.

\subsection*{Acknowledgements}
We are grateful to Alexandre Belin, Vishnu Jejjala, Arjun Kar, Guram Kartvelishvili, Lampros Lamprou, Juan Maldacena, Onkar Parrikar, Simon Ross, and Tomonori Ugajin for useful conversations. MD is supported by the National Science Foundation Graduate Research Fellowship under Grant No. DGE-1845298.  The research of VB, MD, and GS was supported in part by the Simons Foundation through the It From Qubit Collaboration (Grant No.~38559), and by the Department of Energy through grants DE-SC0013528, and QuantISED  DE-SC0020360.  VB also thanks the Aspen Center for Physics, which is supported by National Science Foundation grant PHY-1607611, for hospitality while this work was in progress.

\begin{appendices}

\section{Perturbative Equations of Motion}\label{sec:perturbEOM}

In this appendix, we derive the perturbative equations of motion \eqref{eq:perturbativeEOMs1} and \eqref{eq:perturbativeEOMs2}. In general, bars will indicate background-order quantities. We begin by variation of the full geometric equation of motion \eqref{eq:geomEOM1}. Since we impose self-duality at all orders,
\begin{align}
    0 = (F + \delta F)\wedge \ast(F+\delta F) \sim F_{\alpha \beta \gamma \delta \epsilon} F^{\alpha \beta \gamma \delta \epsilon} \, ,
\end{align}
where the last expression is to all orders. Consequently, the variation of \eqref{eq:geomEOM1} is
\begin{align}
    \delta R_{\mu \nu} -\frac12 \delta R \bar{g}_{\mu \nu} &=  \frac{1}{4\cdot 4!} (\delta F_{\mu \alpha \beta \gamma \delta} \bar{F}_{\nu}^{\:\:\alpha \beta\gamma\delta} + \bar{F}_{\mu \alpha \beta \gamma \delta} \delta F_{\nu}^{\:\:\alpha \beta \gamma \delta} ) \, .
\end{align}
$\bar{R} = 0$ at background order, but it is not obvious that $\delta R = 0$ perturbatively, so we have retained this term for now. Now we can constrain $\delta R$ by tracing both sides, noting that $\bar{g}^{\mu \nu} \delta R_{\mu \nu} = \delta R - \bar{R}_{\mu \nu} \delta g^{\mu \nu}$,
\begin{align}
    -4\delta R - \bar{R}_{\mu \nu} \delta g^{\mu \nu} 
    = \frac{1}{4\cdot 4!} \bar{g}^{\mu \nu} (\delta F_{\mu \alpha \beta \gamma \delta} \bar{F}_{\nu}^{\:\:\alpha \beta\gamma\delta} + \bar{F}_{\mu \alpha \beta \gamma \delta} \delta F_{\nu}^{\:\:\alpha \beta \gamma \delta} ) \, .
\end{align}
Substituting $\bar{R}_{\mu \nu}$ with its background equation of motion and rearranging for $\delta R$ one finds:
\begin{align}
    \delta R = \frac{1}{16\cdot 4!} \delta \left(g^{\mu \nu} F_{\mu \alpha \beta \gamma \delta} F_{\nu}^{\:\:\alpha\beta\gamma\delta} \right) = 0 \, .
\end{align}
That is, $\delta R$ vanishes to all orders as a consequence of the self-duality constraint. The perturbative equation of motion for the metric is therefore simply
\begin{align}
    \delta R_{\mu \nu}&=  \frac{1}{4\cdot 4!} (\delta F_{\mu \alpha \beta \gamma \delta} \bar{F}_{\nu}^{\:\:\alpha \beta\gamma\delta} + \bar{F}_{\mu \alpha \beta \gamma \delta} \delta F_{\nu}^{\:\:\alpha \beta \gamma \delta} ) \, .
\end{align}
    
We now make use of the formula for $\delta R_{\mu \nu}$ in terms of the metric perturbation $\delta g_{\mu \nu} = h_{\mu \nu}$, to first order in the perturbation, to arrive at \eqref{eq:perturbativeEOMs1}:
\begin{align}
\delta R_{\mu \nu} = \nabla_{\lambda} \nabla_{(\mu}h_{\nu)}^{\lambda} - \frac12 \nabla_{\mu} \partial_{\nu} h - \frac12 \nabla_{\lambda} \nabla^{\lambda} h_{\mu \nu} \, ,
\end{align}
where the covariant derivative is taken with respect to $\bar{g}$. We review the derivation of this formula below. The Ricci tensor is
\begin{align}
R_{bd} = R^a_{bad} = \partial_a \Gamma^a_{bd} - \partial_d \Gamma^a_{ba} + \Gamma^s_{bd} \Gamma^a_{sa} - \Gamma^s_{ba} \Gamma^a_{sd}  \, .
\end{align}
Relabeling indices and varying each term independently gives a formula in terms of the variation $\delta \Gamma$
\begin{align}
\begin{aligned}
\delta R_{ab} &=  \partial_c \delta \Gamma^c_{ab} - \partial_b \delta \Gamma^c_{ac} + \delta(\Gamma^s_{ab} \Gamma^c_{sc}) - \delta(\Gamma^s_{ac} \Gamma^c_{sb} )  \\
&= \partial_c \delta \Gamma^c_{ab} - \partial_b \delta \Gamma^c_{ac} + \delta\Gamma^s_{ab} \Gamma^c_{sc} +\Gamma^s_{ab}\delta \Gamma^c_{sc} - \delta\Gamma^s_{ac} \Gamma^c_{sb} -\Gamma^s_{ac}\delta \Gamma^c_{sb} \\
&= \nabla_c (\delta \Gamma^c_{ab})  - \nabla_b (\delta \Gamma^c_{ac} ) \, .
\end{aligned}
\end{align}
To compute the variation of the Christoffel symbols, expand the covariant derivative of the metric perturbations
\begin{align}
\begin{aligned}
\nabla_a h_{bc} &= \nabla_a (\delta g_{bc}) =  \partial_a (\delta g_{bc}) - \Gamma^s_{ab} \delta g_{sc} - \Gamma^s_{ac} \delta g_{bs} \\
&=  \delta  (\partial_a g_{bc}) - \delta(\Gamma^s_{ab}g_{sc}) + \delta \Gamma^s_{ab}g_{sc} - \delta ( \Gamma^s_{ac}  g_{bs} )+ \delta\Gamma^s_{ac}  g_{bs} \\
&= \delta  (\partial_a g_{bc} - \Gamma^s_{ab}g_{sc}- \Gamma^s_{ac}  g_{bs} )+ \delta \Gamma^s_{ab}g_{sc} + \delta\Gamma^s_{ac}  g_{bs} \\
&= \delta  (\nabla_a g_{bc})+ \delta \Gamma^s_{ab}g_{sc} + \delta\Gamma^s_{ac}  g_{bs} \\
&= \delta \Gamma^s_{ab}g_{sc} + \delta\Gamma^s_{ac}  g_{bs} \, ,
\end{aligned}
\end{align}
using metric compatibility. Now cyclically permuting and adding a convenient sign gives
\begin{align}
\begin{aligned}
\nabla_a h_{bc} +  \nabla_b h_{ca} - \nabla_c h_{ab} &=  \delta \Gamma^s_{ab}g_{sc} + \delta\Gamma^s_{ac}  g_{bs} +  \delta \Gamma^s_{bc}g_{sa} + \delta\Gamma^s_{ba}  g_{cs} -  \delta \Gamma^s_{ca}g_{sb} - \delta\Gamma^s_{cb}  g_{as} \\
&= 2\delta \Gamma^s_{ab}g_{sc}   \, .
\end{aligned}
\end{align}
Rearranging and permuting the indices gives
\begin{align}
\delta \Gamma^a_{bc} = \frac12 (\nabla_b h_{c}^a +  \nabla_c h_{b}^a - \nabla^a h_{bc}) \, .
\end{align}
Expanding the variation $\delta R_{ab}$ with this formula, one finds
\begin{align}
\begin{aligned}
\delta R_{ab} &= \nabla_c (\delta \Gamma^c_{ab})  - \nabla_b (\delta \Gamma^c_{ac} ) \\
&= \frac12 \nabla_c (\nabla_a h^c_b + \nabla_b h^c_a - \nabla^c h_{ab})- \frac12 \nabla_b (\nabla_a h + \nabla_c h^c_a - \nabla^c h_{ac}) \\
&= \nabla_c \nabla_{(a} h_{b)}^c - \frac12 \nabla^2 h_{ab} -\frac12 \nabla_a \partial_b h \, ,
\end{aligned}
\end{align}
which was the claimed formula for the variation of the Ricci tensor.

Now we must consider the variation of Maxwell's equations:
\begin{align}
\delta \left( \partial_{\mu} (\sqrt{-g} F^{ \mu \nu \rho \sigma\tau})\right) &= \partial_{\mu} (\delta \sqrt{-g} \bar{F}^{\mu \nu \rho \sigma\tau} + \sqrt{-\bar{g}} \delta  F^{ \mu \nu \rho \sigma\tau}) \, .
\end{align}
Recall the variation of $\sqrt{-g}$, from Sylvester's formula:
\begin{align}
\delta \sqrt{-g} = -\frac12 \sqrt{-\bar{g}} \bar{g}_{\mu \nu} \delta g^{\mu \nu} = \sqrt{-\bar{g}} \frac{h}{2} \, .
\end{align}
Therefore, we find \eqref{eq:perturbativeEOMs2}:
\begin{align}
\partial_{\mu} (\sqrt{-\bar{g}} (\frac{h}{2} \bar{F}^{ \mu \nu \rho \sigma\tau} +  \delta  F^{ \mu \nu \rho \sigma\tau})) = 0 \, .
\end{align}

Lastly, the perturbation to the five-form must leave it to be self-dual. However, one must be careful because the Hodge dual involves factors of the metric that also contribute perturbatively. Let us assume the metric is diagonal and that the only nonzero independent components of the five-form are $F_{t123r}$ and $F_{\theta_1 \ldots \theta_5}$. In terms of the components of the metric and five-form the constraint can be written explicitly as
\begin{align}
    F_{t123r} + \delta F_{t123r} &= \sqrt{-g} g^{\theta_1 \theta_1} \ldots g^{\theta_5 \theta_5} (F_{\theta_1 \ldots \theta_5} + \delta F_{\theta_1 \ldots \theta_5}) \\
    F_{\theta_1 \ldots \theta_5} + \delta F_{\theta_1 \ldots \theta_5} &= -\sqrt{-g} g^{tt} g^{11} g^{22} g^{33} g^{rr} (F_{t123r} + \delta F_{t123r}) \, .
\end{align}
Now removing the background-order equations and expanding perturbatively, we find
\begin{align}
     \delta F_{t123r} &= \sqrt{-\bar{g}} \bar{g}^{\theta_1 \theta_1} \ldots \bar{g}^{\theta_5 \theta_5} \delta F_{\theta_1 \ldots \theta_5} - \sqrt{-\bar{g}} (h^{\theta_1 \theta_1} \ldots \bar{g}^{\theta_5 \theta_5} + \ldots + \bar{g}^{\theta_1 \theta_1} \ldots h^{\theta_5 \theta_5} ) \bar{F}_{\theta_1 \ldots \theta_5}  + \frac{h}{2}  \bar{F}_{t123r} \label{eq:selfdual1}\\
   \delta F_{\theta_1 \ldots \theta_5} &= -\sqrt{-\bar{g}} \bar{g}^{tt} \bar{g}^{11} \bar{g}^{22} \bar{g}^{33} \bar{g}^{rr} \delta F_{t123r} + \sqrt{-\bar{g}} (h^{tt} \bar{g}^{11} \bar{g}^{22} \bar{g}^{33} \bar{g}^{rr} + \ldots +\bar{g}^{tt} \bar{g}^{11} \bar{g}^{22} \bar{g}^{33} h^{rr} ) F_{t123r}  + \frac{h}{2} \bar{F}_{\theta_1 \ldots \theta_5} \, .\label{eq:selfdual2}
\end{align}
To proceed further, one requires more details about the background metric and five-form of interest.

\section{Solving the Linearized Equations} \label{sec:linearizedApp}

In this appendix, we demonstrate the procedure to solve the perturbative equations of motion \eqref{eq:perturbativeEOMs1} and \eqref{eq:perturbativeEOMs2} by hand in the linearized regime, region II. In this regime the ansatz for the metric and five-form perturbations takes the form
\begin{align}
\frac{1}{\alpha'} ds^2 &= L^2 \biggl[-\tilde{r}^2\left(1 + \delta g_{\tilde{t}\tilde{t}} \right) d\tilde{t}^2 +\tilde{r}^2\left(1 + \delta g_{\tilde{i}\tilde{i}}\right) d\vec{\tilde{x}}^2  + \frac{1}{\tilde{r}^2}\left(1+\delta g_{\tilde{r}\tilde{r}}\right)d\tilde{r}^2 +  (1+\delta g_{\Omega \Omega}) d\Omega_5^2 \biggr] \\
\frac{1}{\alpha^{'2}} F &= 4L^4 \biggl[\tilde{r}^3 \left(1 +a(\tilde{r}) \right)  d\tilde{t} \wedge d\tilde{x}^1 \wedge d\tilde{x}^2 \wedge d\tilde{x}^3 \wedge d\tilde{r} \nonumber  \\&\qquad\qquad\qquad\qquad\qquad+ (1+b(\tilde{r}))  \sin^4 \theta_1 \sin^3 \theta_2 \sin^2 \theta_3 \sin \theta_4 d\theta_1 \wedge d\theta_2 \wedge d\theta_3 \wedge d\theta_4 \wedge d\theta_5 \biggr] \, ,
\end{align}
where all metric perturbations are functions only of $\tilde{r}$. To begin, we evaluate the self-duality constraint equations \eqref{eq:selfdual1} and \eqref{eq:selfdual2} on the background metric of vacuum AdS$_5 \times S_5$. They reduce to only one independent equation,
\begin{align}
    2 a-3  \delta g_{\tilde{i}\tilde{i}}- \delta g_{\tilde{t}\tilde{t}}+5  \delta g_{\Omega \Omega }- \delta g_{\tilde{r}\tilde{r}} &= 2b \, ,
\end{align}
identical to \eqref{eq:selfdualBB}. Plugging into Maxwell's equations we also find only one independent equation
\begin{align}
      2  a'-3  \delta g_{\tilde{i}\tilde{i}}'- \delta g_{\tilde{t}\tilde{t}}'+5  \delta g_{\Omega \Omega}'- \delta g_{\tilde{r}\tilde{r}}' &= 0 \, ,
\end{align}
which is identical to \eqref{eq:maxwellBB}. Combining Maxwell's equations and self-duality gives $b' = 0$, meaning that the five-form charge is conserved. The independent geometric equations of motion are
\begin{align}
    -8 (b  - 
    2\delta g_{\Omega\Omega } ) - \frac {1} {2} \tilde{r}\left (\tilde{r} g_{\Omega\Omega }''  + 
    5g_{\Omega\Omega }'  \right) &= 0 \label{eq:geom1}\\
    -16  a+\tilde{r}^2 \delta g_{\tilde{t}\tilde{t}} ''+3 \tilde{r} \delta g_{\tilde{i}\tilde{i}} '+6 \tilde{r} \delta g_{\tilde{t}\tilde{t}}'+5 \tilde{r} \delta g_{\Omega \Omega}'- \tilde{r} \delta g_{\tilde{r}\tilde{r}}'+24  \delta g_{\tilde{i}\tilde{i}} +8  \delta g_{\tilde{t}\tilde{t}} &= 0\label{eq:geom2}\\
    -16  a+\tilde{r}^2 \delta g_{\tilde{i}\tilde{i}}''+8 \tilde{r}\delta g_{\tilde{i}\tilde{i}}'+\tilde{r} \delta g_{\tilde{t}\tilde{t}}'+5 \tilde{r} \delta g_{\Omega \Omega }'- \tilde{r} \delta g_{\tilde{r}\tilde{r}}'+24 \delta g_{\tilde{i}\tilde{i}}+8 \delta g_{\tilde{t}\tilde{t}} &= 0 \label{eq:geom3}\\
    -16  a +3 \tilde{r}^2 \delta g_{\tilde{i}\tilde{i}}''+\tilde{r}^2 \delta g_{\tilde{t}\tilde{t}}''+5 \tilde{r}^2 \delta g_{\Omega \Omega}''+9 \tilde{r}\delta g_{\tilde{i}\tilde{i}}'
    +3 \tilde{r}\delta g_{\tilde{t}\tilde{t}}'\qquad\qquad\qquad\qquad\qquad&\nonumber \\+5 \tilde{r} \delta g_{\Omega \Omega }'-4 \tilde{r} \delta g_{\tilde{r}\tilde{r}}'+24 \delta g_{\tilde{i}\tilde{i}}+8  \delta g_{\tilde{t}\tilde{t}} &= 0 \, .\label{eq:geom4}
\end{align}
Notice that the equation of motion \eqref{eq:geom1} for $\delta g_{\Omega \Omega}$ is independent of the others and may be solved directly, yielding:
\begin{align}
    \delta g_{\Omega \Omega} = \frac{b}{2} + a_1 \tilde{r}^4 + a_2/\tilde{r}^8 \, ,
\end{align}
for constants $a_1, a_2$. Now examine the three remaining geometric equations of motion. By taking the linear combination of \eqref{eq:geom2} + 3\eqref{eq:geom3} - \eqref{eq:geom4}, we find an equation determining $\delta g_{\tilde{r}\tilde{r}}$ in terms of the others:
\begin{align}
    4\delta g_{\tilde{r}\tilde{r}} = 2b + 3\tilde{r}\delta g^{'}_{\tilde{i}\tilde{i}}+ \tilde{r}\delta g^{'}_{\tilde{t}\tilde{t}} +20 (a_1 \tilde{r}^4 -3 a_2 / \tilde{r}^8) \, . \label{eq:rreqn}
\end{align}
Plugging this back into all three equations we find that all three are solved as long as:
\begin{align}
    5(\delta g_{\tilde{i}\tilde{i}}^{'} - \delta g_{\tilde{t}\tilde{t}}^{'}) + r(\delta g_{\tilde{i}\tilde{i}}^{''} - \delta g_{\tilde{t}\tilde{t}}^{''}) = 0 \, .
\end{align}
This is a differential equation in $f (\tilde{r}) = \delta g_{\tilde{i}\tilde{i}} - \delta g_{\tilde{t}\tilde{t}}$ which is solved by $f(\tilde{r}) = c_2 - \frac{c_1}{4\tilde{r}^4}$ where $c_1, c_2$ are constants. Therefore we can relate $\delta g_{\tilde{i}\tilde{i}}$ to $\delta g_{\tilde{t}\tilde{t}}$ via
\begin{align}
    \delta g_{\tilde{i}\tilde{i}} = c_2 - \frac{c_1}{4\tilde{r}^4}+\delta g_{\tilde{t}\tilde{t}} \, . \label{eq:iieqn}
\end{align}
Plugging \eqref{eq:iieqn} into \eqref{eq:rreqn} reduces it to
\begin{align}
    \delta g_{\tilde{r}\tilde{r}} = \frac{b}{2} + \tilde{r}\delta g_{\tilde{t}\tilde{t}}^{'} + \frac34\frac{c_1}{\tilde{r}^4} - 15 \frac{a_2}{\tilde{r}^8} + 5a_1 \tilde{r}^4 \, .
\end{align}
We have consequently fixed the general perturbative solution in the linearized regime in terms of one arbitrary function $\delta g_{\tilde{t}\tilde{t}}$ and five constants $a_1, a_2, c_1, c_2, b$:
\begin{align}
\frac{1}{\alpha'} ds^2 &= L^2 \biggl[-\tilde{r}^2\left(1 + \delta g_{\tilde{t}\tilde{t}} \right) d\tilde{t}^2 +\tilde{r}^2\left(1 + c_2 - \frac{c_1}{4\tilde{r}^4}+\delta g_{\tilde{t}\tilde{t}}\right) d\vec{\tilde{x}}^2  \nonumber\\
&+ \frac{1}{\tilde{r}^2}\left(1+\frac{b}{2} + \tilde{r}\delta g'_{\tilde{t}\tilde{t}} + \frac34\frac{c_1}{\tilde{r}^4} - 15 \frac{a_2}{\tilde{r}^8} + 5a_1 \tilde{r}^4\right)d\tilde{r}^2 +  (1+ \frac{b}{2} + a_1 \tilde{r}^4 + \frac{a_2}{\tilde{r}^8}) d\Omega_5^2 \biggr] \\
\frac{1}{\alpha^{'2}} F &= 4L^4 \biggl[\tilde{r}^3 \left(1 +\frac32 c_2 - 10 \frac{a_2}{\tilde{r}^8} + 2 \delta g_{\tilde{t}\tilde{t}} + \frac{\tilde{r}}{2} \delta g'_{\tilde{t}\tilde{t}}\right)  d\tilde{t} \wedge d\tilde{x}^1 \wedge d\tilde{x}^2 \wedge d\tilde{x}^3 \wedge d\tilde{r} \nonumber  \\&\qquad\qquad\qquad\qquad\qquad+ (1+b)  \sin^4 \theta_1 \sin^3 \theta_2 \sin^2 \theta_3 \sin \theta_4 d\theta_1 \wedge d\theta_2 \wedge d\theta_3 \wedge d\theta_4 \wedge d\theta_5 \biggr] \, .
\end{align}
To fix $\delta g_{\tilde{t}\tilde{t}}$ and the five constants, we compare the solution to the linearized expansion of the solution \eqref{eq:metricnonpert}-\eqref{eq:fiveformnonpert}. A consistent solution is found by taking $a_1 = \epsilon^4 / 2$, $a_2 = c_2 = 0$, $c_1 = -4$, $b=0$, and 
\begin{align}
    \delta g_{\tilde{t}\tilde{t}} = -\frac{1}{\tilde{r}^4} - \frac12 (\epsilon \tilde{r})^4 \, ,
\end{align}
leading to the solution \eqref{eq:linear1}-\eqref{eq:linear2}.

\end{appendices}

\bibliographystyle{JHEP}
\bibliography{Wormholebib}

\providecommand{\href}[2]{#2}\begingroup\raggedright\begin{thebibliography}{10}

\bibitem{vanRaams}
M.~Van~Raamsdonk, \emph{Building up spacetime with quantum entanglement},
  \href{http://dx.doi.org/https://doi.org/10.1007/s10714-010-1034-0}{\emph{General
  Relativity and Gravitation} {\bf 42} (June, 2010) 2323 -- 2329},
  [\href{https://arxiv.org/abs/arXiv:1005.3035}{{\tt arXiv:1005.3035}}].

\bibitem{Maldacena_2003}
J.~Maldacena, \emph{Eternal black holes in anti-de sitter},
  \href{http://dx.doi.org/https://doi.org/10.1088/1126-6708/2003/04/021}{\emph{JHEP}
  {\bf 2003} (apr, 2003) 021 -- 021},
  [\href{https://arxiv.org/abs/arXiv:hep-th/0106112}{{\tt
  arXiv:hep-th/0106112}}].

\bibitem{Balasubramanian_2014}
V.~Balasubramanian, P.~Hayden, A.~Maloney, D.~Marolf and S.~F. Ross,
  \emph{Multiboundary wormholes and holographic entanglement},
  \href{http://dx.doi.org/https://doi.org/10.1088/0264-9381/31/18/185015}{\emph{Classical
  and Quantum Gravity} {\bf 31} (sep, 2014) 185015},
  [\href{https://arxiv.org/abs/arXiv:1406.2663}{{\tt arXiv:1406.2663}}].

\bibitem{Marolf_2015}
D.~Marolf, H.~Maxfield, A.~Peach and S.~Ross, \emph{Hot multiboundary wormholes
  from bipartite entanglement},
  \href{http://dx.doi.org/https://doi.org/10.1088/0264-9381/32/21/215006}{\emph{Classical
  and Quantum Gravity} {\bf 32} (oct, 2015) 215006},
  [\href{https://arxiv.org/abs/arXiv:1506.04128}{{\tt arXiv:1506.04128}}].

\bibitem{doi:10.1002/prop.201300020}
J.~Maldacena and L.~Susskind, \emph{Cool horizons for entangled black holes},
  \href{http://dx.doi.org/10.1002/prop.201300020}{\emph{Fortschritte der
  Physik} {\bf 61} (2013) 781--811},
  [\href{https://arxiv.org/abs/arXiv:1306.0533}{{\tt arXiv:1306.0533}}].

\bibitem{KLEBANOV199989}
I.~R. Klebanov and E.~Witten, \emph{Ads/cft correspondence and symmetry
  breaking},
  \href{http://dx.doi.org/https://doi.org/10.1016/S0550-3213(99)00387-9}{\emph{Nuclear
  Physics B} {\bf 556} (1999) 89 -- 114},
  [\href{https://arxiv.org/abs/arXiv:hep-th/9905104}{{\tt
  arXiv:hep-th/9905104}}].

\bibitem{PhysRevD.62.086003}
M.~Cveti\ifmmode~\check{c}\else \v{c}\fi{}, S.~S. Gubser, H.~L\"u and C.~N.
  Pope, \emph{Symmetric potentials of gauged supergravities in diverse
  dimensions and coulomb branch of gauge theories},
  \href{http://dx.doi.org/10.1103/PhysRevD.62.086003}{\emph{Phys. Rev. D} {\bf
  62} (Sep, 2000) 086003},
  [\href{https://arxiv.org/abs/arXiv:hep-th/9909121}{{\tt
  arXiv:hep-th/9909121}}].

\bibitem{CoulombBranch}
Y.-Y. Wu, \emph{A note on ads/sym correspondence on the coulomb branch},
  \href{https://arxiv.org/abs/arXiv:hep-th/9809055}{{\tt
  arXiv:hep-th/9809055}}.

\bibitem{PhysRevD.60.127902}
A.~Hashimoto, \emph{Holographic description of d3-branes in flat space},
  \href{http://dx.doi.org/10.1103/PhysRevD.60.127902}{\emph{Phys. Rev. D} {\bf
  60} (Nov, 1999) 127902},
  [\href{https://arxiv.org/abs/arXiv:hep-th/9903227}{{\tt
  arXiv:hep-th/9903227}}].

\bibitem{DUFF1991409}
M.~Duff and J.~Lu, \emph{The self-dual type iib superthreebrane},
  \href{http://dx.doi.org/https://doi.org/10.1016/0370-2693(91)90290-7}{\emph{Physics
  Letters B} {\bf 273} (1991) 409 -- 414}.

\bibitem{HOROWITZ1991197}
G.~T. Horowitz and A.~Strominger, \emph{Black strings and p-branes},
  \href{http://dx.doi.org/https://doi.org/10.1016/0550-3213(91)90440-9}{\emph{Nuclear
  Physics B} {\bf 360} (1991) 197 -- 209}.

\bibitem{largeN}
J.~Maldacena, \emph{The large-n limit of superconformal field theories and
  supergravity},
  \href{http://dx.doi.org/https://doi.org/10.1023/A:1026654312961}{\emph{International
  Journal of Theoretical Physics} (apr, 1999) },
  [\href{https://arxiv.org/abs/arXiv:hep-th/9711200}{{\tt
  arXiv:hep-th/9711200}}].

\bibitem{Kraus_1999}
P.~Kraus, F.~Larsen and S.~P. Trivedi, \emph{The coulomb branch of gauge theory
  from rotating branes},
  \href{http://dx.doi.org/10.1088/1126-6708/1999/03/003}{\emph{Journal of High
  Energy Physics} {\bf 1999} (mar, 1999) 003--003},
  [\href{https://arxiv.org/abs/arXiv:hep-th/9811120}{{\tt
  arXiv:hep-th/9811120}}].

\bibitem{NAYEK2017192}
K.~Nayek and S.~Roy, \emph{Decoupling limit and throat geometry of non-susy d3
  brane},
  \href{http://dx.doi.org/https://doi.org/10.1016/j.physletb.2017.01.007}{\emph{Physics
  Letters B} {\bf 766} (2017) 192 -- 195},
  [\href{https://arxiv.org/abs/arXiv:1608.05036}{{\tt arXiv:1608.05036}}].

\bibitem{BERGMAN2009300}
A.~Bergman, H.~Lü, J.~Mei and C.~Pope, \emph{Ads wormholes},
  \href{http://dx.doi.org/https://doi.org/10.1016/j.nuclphysb.2008.11.008}{\emph{Nuclear
  Physics B} {\bf 810} (2009) 300 -- 315},
  [\href{https://arxiv.org/abs/arXiv:0808.2481}{{\tt arXiv:0808.2481}}].

\bibitem{Maldacena_2004}
J.~Maldacena and L.~Maoz, \emph{Wormholes in {AdS}},
  \href{http://dx.doi.org/https://doi.org/10.1088/1126-6708/2004/02/053}{\emph{Journal
  of High Energy Physics} {\bf 2004} (feb, 2004) 053--053},
  [\href{https://arxiv.org/abs/arXiv:hep-th/0401024}{{\tt
  arXiv:hep-th/0401024}}].

\bibitem{susywormhole}
A.~Anabalon, B.~{de Wit} and J.~Oliva, \emph{Supersymmetric traversable
  wormholes},
  \href{http://dx.doi.org/https://doi.org/10.1007/JHEP09(2020)109}{\emph{JHEP}
  (aug, 2020) }, [\href{https://arxiv.org/abs/arXiv:2001.00606}{{\tt
  arXiv:2001.00606}}].

\bibitem{costa}
M.~S. Costa, \emph{Absorption by double-centered d3-branes and the coulomb
  branch of $\mathcal{N} = 4$ {SYM} theory},
  \href{http://dx.doi.org/10.1088/1126-6708/2000/05/041}{\emph{Journal of High
  Energy Physics} {\bf 2000} (may, 2000) 041--041},
  [\href{https://arxiv.org/abs/arXiv:hep-th/9912073}{{\tt
  arXiv:hep-th/9912073}}].

\bibitem{Michelson_1999}
J.~Michelson and A.~Strominger, \emph{Superconformal multi-black hole quantum
  mechanics},
  \href{http://dx.doi.org/10.1088/1126-6708/1999/09/005}{\emph{Journal of High
  Energy Physics} {\bf 1999} (sep, 1999) 005--005},
  [\href{https://arxiv.org/abs/arXiv:hep-th/9908044}{{\tt
  arXiv:hep-th/9908044}}].

\bibitem{GALLOWAY2001255}
G.~Galloway, K.~Schleich, D.~Witt and E.~Woolgar, \emph{The ads/cft
  correspondence and topological censorship},
  \href{http://dx.doi.org/https://doi.org/10.1016/S0370-2693(01)00335-5}{\emph{Physics
  Letters B} {\bf 505} (2001) 255 -- 262},
  [\href{https://arxiv.org/abs/arXiv:hep-th/9912119}{{\tt
  arXiv:hep-th/9912119}}].

\bibitem{Morris:1988tu}
M.~Morris, K.~Thorne and U.~Yurtsever, \emph{{Wormholes, Time Machines, and the
  Weak Energy Condition}},
  \href{http://dx.doi.org/10.1103/PhysRevLett.61.1446}{\emph{Phys. Rev. Lett.}
  {\bf 61} (1988) 1446--1449}.

\bibitem{PhysRevLett.90.201102}
M.~Visser, S.~Kar and N.~Dadhich, \emph{Traversable wormholes with arbitrarily
  small energy condition violations},
  \href{http://dx.doi.org/10.1103/PhysRevLett.90.201102}{\emph{Phys. Rev.
  Lett.} {\bf 90} (May, 2003) 201102},
  [\href{https://arxiv.org/abs/arXiv:gr-qc/0301003}{{\tt
  arXiv:gr-qc/0301003}}].

\bibitem{PhysRevLett.61.1446}
M.~S. Morris, K.~S. Thorne and U.~Yurtsever, \emph{Wormholes, time machines,
  and the weak energy condition},
  \href{http://dx.doi.org/10.1103/PhysRevLett.61.1446}{\emph{Phys. Rev. Lett.}
  {\bf 61} (Sep, 1988) 1446--1449}.

\bibitem{PhysRevLett.81.746}
D.~Hochberg and M.~Visser, \emph{Null energy condition in dynamic wormholes},
  \href{http://dx.doi.org/10.1103/PhysRevLett.81.746}{\emph{Phys. Rev. Lett.}
  {\bf 81} (Jul, 1998) 746--749}.

\bibitem{GaoJafferisWall}
P.~Gao, D.~L. Jafferis and A.~C. Wall, \emph{Traversable wormholes via a double
  trace deformation},
  \href{http://dx.doi.org/https://doi.org/10.1007/JHEP12(2017)151}{\emph{JHEP}
  (dec, 2017) }, [\href{https://arxiv.org/abs/arXiv:1608.05687}{{\tt
  arXiv:1608.05687}}].

\bibitem{GJWBounds}
B.~Freivogel, D.~A. Galante, D.~Nikolakopolou and A.~Rotundo, \emph{Traversable
  wormholes in ads and bounds on information transfer},
  \href{http://dx.doi.org/https://doi.org/10.1007/JHEP01(2020)050}{\emph{JHEP}
  (jan, 2020) }, [\href{https://arxiv.org/abs/arXiv:1907.13140}{{\tt
  arXiv:1907.13140}}].

\bibitem{Fu_2019}
Z.~Fu, B.~Grado-White and D.~Marolf, \emph{A perturbative perspective on
  self-supporting wormholes},
  \href{http://dx.doi.org/https://doi.org/10.1088/1361-6382/aafcea}{\emph{Classical
  and Quantum Gravity} {\bf 36} (jan, 2019) 045006},
  [\href{https://arxiv.org/abs/arXiv:1807.07917}{{\tt arXiv:1807.07917}}].

\bibitem{Fu_2019b}
D.~Marolf and S.~McBride, \emph{Simple perturbatively traversable wormholes
  from bulk fermions},
  \href{http://dx.doi.org/https://doi.org/10.1007/JHEP11(2019)037}{\emph{JHEP}
  (oct, 2019) }, [\href{https://arxiv.org/abs/arXiv:1908.03998}{{\tt
  arXiv:1908.03998}}].

\bibitem{SM_wormhole}
J.~Maldacena, A.~Milekhin and F.~Popov, \emph{Traversable wormholes in four
  dimensions},  \href{https://arxiv.org/abs/arXiv:1807.04726}{{\tt
  arXiv:1807.04726}}.

\bibitem{selfsupp}
A.~Anand and P.~K. Tripathy, \emph{Self-supporting wormholes with massive
  vector field},  \href{https://arxiv.org/abs/arXiv:2008.10920}{{\tt
  arXiv:2008.10920}}.

\bibitem{Horowitz_2019}
G.~T. Horowitz, D.~Marolf, J.~E. Santos and D.~Wang, \emph{Creating a
  traversable wormhole},
  \href{http://dx.doi.org/https://doi.org/10.1088/1361-6382/ab436f}{\emph{Classical
  and Quantum Gravity} {\bf 36} (sep, 2019) 205011},
  [\href{https://arxiv.org/abs/arXiv:1904.02187}{{\tt arXiv:1904.02187}}].

\bibitem{Fu_2019a}
Z.~Fu, B.~Grado-White and D.~Marolf, \emph{Traversable asymptotically flat
  wormholes with short transit times},
  \href{http://dx.doi.org/https://doi.org/10.1088/1361-6382/ab56e4}{\emph{Classical
  and Quantum Gravity} {\bf 36} (nov, 2019) 245018},
  [\href{https://arxiv.org/abs/arXiv:1908.03273}{{\tt arXiv:1908.03273}}].

\bibitem{doi:10.1002/prop.201700034}
J.~Maldacena, D.~Stanford and Z.~Yang, \emph{Diving into traversable
  wormholes},
  \href{http://dx.doi.org/10.1002/prop.201700034}{\emph{Fortschritte der
  Physik} {\bf 65} (2017) 1700034},
  [\href{https://arxiv.org/abs/arXiv:1704.05333}{{\tt arXiv:1704.05333}}].

\bibitem{SYKwormhole}
J.~Maldacena and X.-L. Qi, \emph{Eternal traversable wormhole},
  \href{https://arxiv.org/abs/arXiv:1804.00491}{{\tt arXiv:1804.00491}}.

\bibitem{GaoJafferisSYK}
P.~Gao and D.~L. Jafferis, \emph{A traversable wormhole teleportation protocol
  in the syk model},  \href{https://arxiv.org/abs/arXiv:1911.07416}{{\tt
  arXiv:1911.07416}}.

\bibitem{SYKwormhole2}
J.~Maldacena and A.~Milekhin, \emph{Syk wormhole formation in real time},
  \href{https://arxiv.org/abs/arXiv:1912.03276}{{\tt arXiv:1912.03276}}.

\bibitem{GJWprobes}
D.~Bak, C.~Kim and S.-H. Yi, \emph{Experimental probes of traversable
  wormholes},
  \href{http://dx.doi.org/https://doi.org/10.1007/JHEP12(2019)005}{\emph{JHEP}
  {\bf 2019} (nov, 2019) }, [\href{https://arxiv.org/abs/arXiv:1907.13465}{{\tt
  arXiv:1907.13465}}].

\bibitem{wormhole_signaling}
A.~R. Brown, H.~Gharibyan, S.~Leichenauer, H.~W. Lin, S.~Nezami, G.~Salton
  et~al., \emph{Quantum gravity in the lab: Teleportation by size and
  traversable wormholes},  \href{https://arxiv.org/abs/arXiv:1911.06314}{{\tt
  arXiv:1911.06314}}.

\bibitem{connectivity}
F.~Aprile and V.~Niarchos, \emph{Large-n transitions of the connectivity
  index},
  \href{http://dx.doi.org/https://doi.org/10.1007/JHEP02(2015)083}{\emph{JHEP}
  (jan, 2015) }, [\href{https://arxiv.org/abs/arXiv:1410.7773}{{\tt
  arXiv:1410.7773}}].

\bibitem{BRINK197777}
L.~Brink, J.~H. Schwarz and J.~Scherk, \emph{Supersymmetric yang-mills
  theories},
  \href{http://dx.doi.org/https://doi.org/10.1016/0550-3213(77)90328-5}{\emph{Nuclear
  Physics B} {\bf 121} (1977) 77 -- 92}.

\bibitem{DouglasTaylor}
M.~R. Douglas and W.~{Taylor IV}, \emph{Branes in the bulk of anti-de sitter
  space},  \href{https://arxiv.org/abs/arXiv:hep-th/9807225}{{\tt
  arXiv:hep-th/9807225}}.

\bibitem{HOOFT1974461}
G.~{'t Hooft}, \emph{A planar diagram theory for strong interactions},
  \href{http://dx.doi.org/https://doi.org/10.1016/0550-3213(74)90154-0}{\emph{Nuclear
  Physics B} {\bf 72} (1974) 461 -- 473}.

\bibitem{PhysRevLett.27.1688}
S.~Weinberg, \emph{Physical processes in a convergent theory of the weak and
  electromagnetic interactions},
  \href{http://dx.doi.org/10.1103/PhysRevLett.27.1688}{\emph{Phys. Rev. Lett.}
  {\bf 27} (Dec, 1971) 1688--1691}.

\bibitem{doi:10.1063/1.1372177}
N.~Drukker and D.~J. Gross, \emph{An exact prediction of n=4 supersymmetric
  yang–mills theory for string theory},
  \href{http://dx.doi.org/10.1063/1.1372177}{\emph{Journal of Mathematical
  Physics} {\bf 42} (2001) 2896--2914},
  [\href{https://arxiv.org/abs/arXiv:hep-th/0010274}{{\tt
  arXiv:hep-th/0010274}}].

\bibitem{pol95}
J.~Polchinski, \emph{Dirichlet branes and ramond-ramond charges},
  \href{http://dx.doi.org/10.1103/PhysRevLett.75.4724}{\emph{Phys. Rev. Lett.}
  {\bf 75} (Dec, 1995) 4724--4727},
  [\href{https://arxiv.org/abs/arXiv:hep-th/9510017}{{\tt
  arXiv:hep-th/9510017}}].

\bibitem{DUFF1995213}
M.~Duff, R.~R. Khuri and J.~Lu, \emph{String solitons},
  \href{http://dx.doi.org/https://doi.org/10.1016/0370-1573(95)00002-X}{\emph{Physics
  Reports} {\bf 259} (1995) 213 -- 326},
  [\href{https://arxiv.org/abs/arXiv:hep-th/9412184}{{\tt
  arXiv:hep-th/9412184}}].

\bibitem{peet}
A.~Peet, \emph{TASI Lectures on Black Holes in String Theory}, pp.~353--433.
\newblock World Scientific, 1999.
\newblock \href{https://arxiv.org/abs/arXiv:hep-th/0008241}{{\tt
  arXiv:hep-th/0008241}}.

\bibitem{kiritsis}
E.~Kiritsis, \emph{String Theory in a Nutshell}.
\newblock Princeton University Press, 2~ed., Apr., 2019.

\bibitem{johnson}
C.~V. Johnson, \emph{D-Branes}.
\newblock Cambridge University Press, 2003.

\bibitem{amerd}
M.~Ammon and J.~Erdmenger, \emph{Gauge/Gravity Duality: Foundations and
  Applications}.
\newblock Cambridge University Press, May, 2015.

\bibitem{Verlinde2020}
H.~Verlinde, \emph{Er = epr revisited: On the entropy of an einstein-rosen
  bridge},  \href{https://arxiv.org/abs/arXiv:2003.13117}{{\tt
  arXiv:2003.13117}}.

\bibitem{GIBBONS1998603}
G.~Gibbons, \emph{Born-infeld particles and dirichlet p-branes},
  \href{http://dx.doi.org/https://doi.org/10.1016/S0550-3213(97)00795-5}{\emph{Nuclear
  Physics B} {\bf 514} (1998) 603 -- 639},
  [\href{https://arxiv.org/abs/arXiv:hep-th/9709027}{{\tt
  arXiv:hep-th/9709027}}].

\bibitem{INTRILIGATOR200099}
K.~Intriligator, \emph{Maximally supersymmetric rg flows and ads duality},
  \href{http://dx.doi.org/https://doi.org/10.1016/S0550-3213(99)00803-2}{\emph{Nuclear
  Physics B} {\bf 580} (2000) 99 -- 120},
  [\href{https://arxiv.org/abs/arXiv:hep-th/9909082}{{\tt
  arXiv:hep-th/9909082}}].

\bibitem{OGwitten}
E.~Witten, \emph{Anti de sitter space and holography},
  \href{http://dx.doi.org/https://dx.doi.org/10.4310/ATMP.1998.v2.n2.a2}{\emph{Advances
  in Theoretical and Mathematical Physics} {\bf 2} (jan, 1998) },
  [\href{https://arxiv.org/abs/arXiv:hep-th/9802150}{{\tt
  arXiv:hep-th/9802150}}].

\bibitem{PhysRevD.59.104021}
V.~Balasubramanian, P.~Kraus, A.~Lawrence and S.~P. Trivedi, \emph{Holographic
  probes of anti--de sitter spacetimes},
  \href{http://dx.doi.org/10.1103/PhysRevD.59.104021}{\emph{Phys. Rev. D} {\bf
  59} (Apr, 1999) 104021},
  [\href{https://arxiv.org/abs/arXiv:hep-th/9808017}{{\tt
  arXiv:hep-th/9808017}}].

\bibitem{Arutyunov:2000ku}
G.~Arutyunov, S.~Frolov and A.~C. Petkou, \emph{{Operator product expansion of
  the lowest weight CPOs in $\mathcal N=4$ SYM$_4$ at strong coupling}},
  \href{http://dx.doi.org/10.1016/S0550-3213(00)00439-9}{\emph{Nucl. Phys. B}
  {\bf 586} (2000) 547--588},
  [\href{https://arxiv.org/abs/arXiv:hep-th/0005182}{{\tt
  arXiv:hep-th/0005182}}].

\bibitem{Chai:2020zgq}
N.~Chai, S.~Chaudhuri, C.~Choi, Z.~Komargodski, E.~Rabinovici and M.~Smolkin,
  \emph{{Thermal Order in Conformal Theories}},
  \href{https://arxiv.org/abs/arXiv:2005.03676}{{\tt arXiv:2005.03676}}.

\bibitem{Myers:1999ps}
R.~C. Myers, \emph{{Dielectric branes}},
  \href{http://dx.doi.org/10.1088/1126-6708/1999/12/022}{\emph{JHEP} {\bf 12}
  (1999) 022}, [\href{https://arxiv.org/abs/arXiv:hep-th/9910053}{{\tt
  arXiv:hep-th/9910053}}].

\bibitem{McGreevy:2000cw}
J.~McGreevy, L.~Susskind and N.~Toumbas, \emph{{Invasion of the giant gravitons
  from Anti-de Sitter space}},
  \href{http://dx.doi.org/10.1088/1126-6708/2000/06/008}{\emph{JHEP} {\bf 06}
  (2000) 008}, [\href{https://arxiv.org/abs/arXiv:hep-th/0003075}{{\tt
  arXiv:hep-th/0003075}}].

\bibitem{Balasubramanian:2001nh}
V.~Balasubramanian, M.~Berkooz, A.~Naqvi and M.~J. Strassler, \emph{{Giant
  gravitons in conformal field theory}},
  \href{http://dx.doi.org/10.1088/1126-6708/2002/04/034}{\emph{JHEP} {\bf 04}
  (2002) 034}, [\href{https://arxiv.org/abs/arXiv:hep-th/0107119}{{\tt
  arXiv:hep-th/0107119}}].

\bibitem{Corley:2001zk}
S.~Corley, A.~Jevicki and S.~Ramgoolam, \emph{{Exact correlators of giant
  gravitons from dual N=4 SYM theory}},
  \href{http://dx.doi.org/10.4310/ATMP.2001.v5.n4.a6}{\emph{Adv. Theor. Math.
  Phys.} {\bf 5} (2002) 809--839},
  [\href{https://arxiv.org/abs/arXiv:hep-th/0111222}{{\tt
  arXiv:hep-th/0111222}}].

\bibitem{Balasubramanian:2004nb}
V.~Balasubramanian, D.~Berenstein, B.~Feng and M.-x. Huang, \emph{{D-branes in
  Yang-Mills theory and emergent gauge symmetry}},
  \href{http://dx.doi.org/10.1088/1126-6708/2005/03/006}{\emph{JHEP} {\bf 03}
  (2005) 006}, [\href{https://arxiv.org/abs/arXiv:hep-th/0411205}{{\tt
  arXiv:hep-th/0411205}}].

\end{thebibliography}\endgroup


\providecommand{\href}[2]{#2}\begingroup\raggedright\endgroup

\end{document}